\documentclass[aps,prd,superscriptaddress,preprintnumbers,nofootinbib,12pt]{revtex4-2}

\usepackage{graphicx,amsmath,amssymb}
\usepackage{hyperref}
\usepackage{bm}
\usepackage{hyphenat}
\usepackage{rotating}

\hypersetup{colorlinks=true, linkcolor = [rgb]{0,0.08,0.45}, citecolor = [rgb]{0,0.08,0.45}, urlcolor = [rgb]{0,0.08,0.45}}

\newcommand{\bra}[1]{\langle#1\vert}
\newcommand{\ket}[1]{\vert#1\rangle}
\newcommand{\req}[1]{(\ref{#1})}
\newcommand{\lb}{\label}

\begin{document}

\preprint{SLAC-PUB-17727}

\title{\LARGE Entropy from entangled parton states \\  \vspace{-2pt} and high-energy scattering behavior \vspace{40pt}}


\author{Hans~G\"{u}nter~Dosch}
\email[]{h.g.dosch@gmail.com}
\affiliation{Institut f\"{u}r Theoretische Physik der Universit\"{a}t,  D-69120 Heidelberg, Germany}

\author{Guy~F.~de~T\'eramond}
\email[]{guy.deteramond@ucr.ac.cr}
\affiliation{Laboratorio de F\'isica Te\'orica y Computacional, Universidad de Costa Rica, 11501 San Jos\'e, Costa Rica}

\author{Stanley~J.~Brodsky}
\email[]{sjbth@slac.stanford.edu}
\affiliation{SLAC National Accelerator Laboratory, Stanford University, Stanford, CA 94309, USA}



\begin{abstract}

\vspace{40pt}

The relation between the gluon density in a hadron and entanglement entropy can shed a new light on the high energy scattering behavior of hadrons. Using the holographic light-front QCD framework the growth above the classical geometric cross section is directly related to the increase of the internal quantum entropy from the entangled parton distribution in hadrons. A rather consistent picture emerges from the scale dependence of the Pomeron from the QCD evolution of the gluon distribution function $g(x, \mu)$, the rising of the integrated cross section in photoproduction of vector mesons,  the deep inelastic scattering (DIS) experiments at HERA, hadron multiplicity and quantum entropy.  We also point out a possible analogy between parton entanglement entropy and the black-hole entropy of Bekenstein and Hawking.

\end{abstract}

\maketitle

\newpage

\section{Introduction}

In their analysis of  entanglement at the subnuclear level, Kharzeev and Levin~\cite{kl1,kl2} have drawn the attention to entropy. The quantum entropy of any bound  state is zero, but  Kharzeev and Levin considered the entropy of the partons resolved in a deep inelastic scattering (DIS) experiment
\begin{align} \lb{sdis1}
S_{DIS} = \ln N(x, Q^2) ,
\end{align}
where $N(x, Q^2)$  is the number of partons in a hadron with longitudinal light-front momentum fraction $x$ of the struck parton in the target hadron and $Q^2 = -q^2 > 0$ is the momentum transfer. The DIS entropy $S_{DIS}$ is the logarithm of the number of degrees of freedom in the DIS measurement. It represents the entropy of entanglement between the  proton components probed by deep inelastic scattering and the rest of the proton: the number of produced final-state spectator quark and gluon partons.  Since the partons cannot be isolated as asymptotic states, the number of partons is not a directly observable quantity, but depends on the virtuality scale  of the process $Q^2$, and therefore the quantum DIS entropy is not directly observable.

The quantum (von-Neumann) entropy of a state described by the density operator $\rho$ is given in analogy to the classical (Boltzmann) entropy by the expectation value of the trace of the statistical  operator
\begin{align} \lb{sq} 
S_Q = - tr [ \rho \,\ln \rho] =-  \sum_i p_i \ln p_i ,
\end{align}
where the $p_i$ are the eigenvalues of $\rho$; they give the probability to find the system in the state $\ket{i}$. A pure state $\ket{\psi}$, like an elementary particle, has therefore the quantum {entropy~0}.  Under the assumption  that a  hadron state $\ket{\psi}$ consists of $N$ interacting constituents and that these partons do not factorize into hadronic substates,  the partons in the hadron are entangled.

In deep inelastic scattering (DIS) a measurement projects only on a single parton $\ket{j}$. The pure state before the measurement is thus, after a measurement, a mixture described by the statistical operator $\rho = \sum_{j=1}^N \, p_j\, \ket{j} \bra{j}$, where $p_j$ is the probability of hitting the state $\ket{j}$; that is, after measurement, which traces out the unobserved components of the state, the entropy is given by the expression \req{sq}.  For very slow partons the gluons are dominant, and we restrain ourselves to this region in order to treat all partons on equal footing. Therefore, we have equal probability for each parton to be hit and thus one expects $p_j= 1/N(x) $, hence  $S_Q = \ln N(x)$,  where $N(x)$ is the number of partons (gluons) with longitudinal momentum fraction $x$, and the result is \req{sdis1}.

In this letter we discuss the consequences of the relation of the DIS entropy introduced in Ref.~\cite{kl1} with the determination of the gluon distribution given in Ref.~\cite{deTeramond:2021lxc} from the high energy scattering of hadrons and holographic light-front QCD~\cite{Brodsky:2006uqa,deTeramond:2008ht,physrep}. This specific connection allows us to relate the increase above the proton-proton geometric cross section to the number of entangled partons in the proton probed in the DIS process. Furthermore, using the light-front holographic framework and the QCD evolution results from Ref.~\cite{pomhol}, we show that the scale dependence of the gluon distribution produces the Pomeron-dominated energy dependence of the DIS cross section. It also accounts for the logarithmic dependence on the observational scale $\mu$,  which, for our purposes, can be identified with the photon virtuality $\mu^2 = Q^2$ in the DIS measurement $\gamma^*(q^2) + p \to X$;  this is an essential prediction of the scale-dependent Pomeron intercept discussed in Ref.~\cite{pomhol}.

In Refs.~\cite{kl1, kl2, KL, Hentschinski:2023izh} it was further assumed that the entropy of the final hadronic state can,  under certain kinematical conditions, be equated to the DIS entropy. This assumption created delicate experimental and theoretical questions, since parton states are not asymptotic states and the final state, after a DIS event,  consists of hadrons which are supposed to be formed after the hadronization of the partons. By examining the hadron multiplicity in the final state, we will briefly discuss how this problem can be also addressed using the concept of a unique scale-dependent Pomeron introduced in Ref.~\cite{DF15}, and recently discussed in Ref.~\cite{pomhol} in the framework of holographic light-front QCD (HLFQCD).

\section{High energy scattering and entanglement entropy}

\subsection{High energy behavior of hadron cross sections}

One of the  most important concepts and tools in high energy scattering is the notion of Regge poles. Originally introduced in scattering in quantum mechanical potential theory by T. Regge~\cite{Reg59}, it was applied to particle physics by Chew and Frautschi~\cite{CF61}. The guiding principle of the latter authors was the notion of duality. This principle  states, roughly speaking, that the same fields which create the interaction between hadrons are  the quantum fields which show up as particles or resonances. It follows from rather general principles of quantum field theory and is therefore  independent of the specific  dynamics.

\begin{figure}[h]
\includegraphics[width=12.6cm]{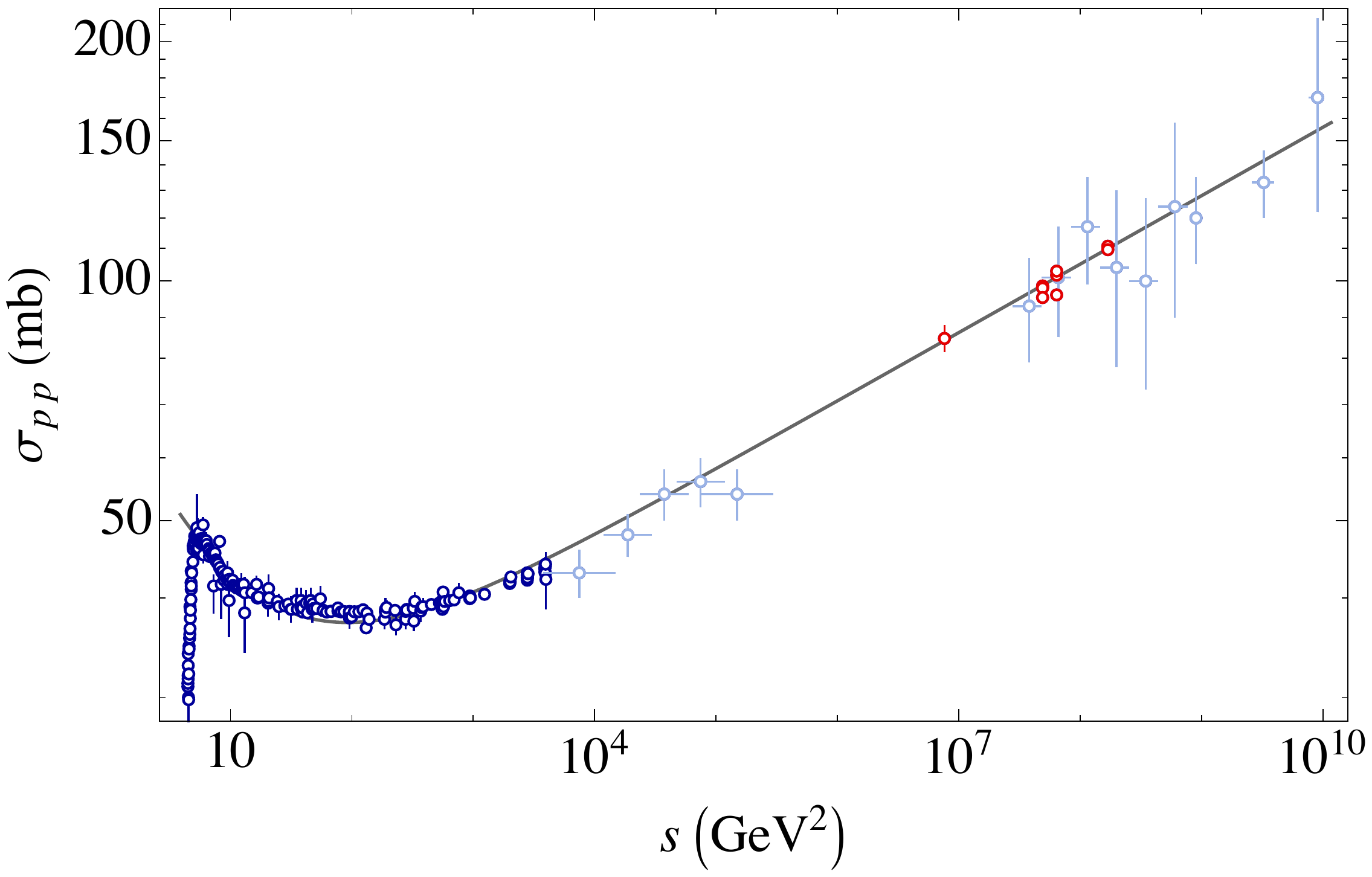}
\caption{\lb{regpp}The total proton-proton scattering cross section as a function of  $s$, the CM energy squared. The straight grey line in the figure shows the energy dependence  $s^{0.086}$, as predicted by a Pomeron with intercept $\alpha_P(0) = 1.086$. The pale blue data points are results from cosmic ray experiments and the red are recent LHC data. Figure adapted  from \cite{PDG20}, Fig. 52.6. }
\end{figure}

In lowest order Regge theory of particle scattering (one pole exchange)  the dependence of  the total cross section on the squared center of mass energy $s $ of the two scattered particles is obtained from the forward scattering amplitude using the optical theorem. Therefore it is determined at $t = 0$ by the intercept of the trajectory in a Chew-Frautschi plot~\cite{CF61}  and the cross section is proportional to $s^{\alpha_P(0)-1}$.   In Fig.~\ref{regpp} a  double logarithmic plot of the proton-proton total cross section {\it vs} the squared center of mass energy $s$  is displayed. One sees that the Pomeron contribution $s^{0.086}$ gives an excellent description of the cross section over 6 orders of magnitude of $s$.  For values of $s \lesssim 10^3$ GeV$^2$ another Regge pole exchange, the $\rho$ trajectory with an intercept 1/2~\cite{deTeramond:2018ecg} gives an additional contribution which vanishes as $s^{-1/2}$ at large $s$. By adding both components 
\begin{align}
\sigma_{pp} =  a (s/s_0)^{0.086} + b (s/s_0)^{-0.5},
\end{align}
with $a = 24 \, \rm{mb}$, $b = 28  \, \rm{mb}$ and $s_0 \equiv 4 m_p^2 $, we find a very good description of the total proton-proton cross section over 9 orders of magnitude in terms of the Pomeron and the $\rho-\omega$ trajectories only.

The Froissart-Martin bound~\cite{Fro61,Mar65} states, however, that the total cross section of two colliding hadrons cannot grow faster than $\ln^2\left(s / s_0\right)$. This result is based on very general assumptions of quantum field theory, especially on the unitarity of the scattering matrix.  Therefore, the very high energy behavior described by single Pomeron exchange has to be modified by unitarity corrections (two and more Pomeron exchanges), but as can be seen from Fig.~\ref{regpp} these corrections would affect only a region well above $10^{10}$ GeV$^2$.

For later discussion we add here a general remark on the high energy behavior: From classical, even wave mechanical, considerations one expects a constant cross section determined by the geometrical extension of the scattered objects. However, in the quantum field theoretical shock-wave model put forward by Heisenberg~\cite{Hei52}, the total cross section increases as $\ln^2 s$ due to the possibility of particle creation. It thus is equal to the upper limit later predicted by the Froissart-Martin bound based on principles of quantum theory.

\subsection{DIS entropy and the Pomeron intercept}

In QCD the gluon distribution at a given virtuality scale $\mu$ is obtained from  DGLAP evolution~\cite{AP77, Dok77, hera15} from the behavior of the DIS cross sections $\gamma^* + p \to X$. The gluon distribution $g(x, \mu)$ is not an observable, and 
consequently, the parton distribution function depends also on the scale $\mu$
 \begin{align} \lb{Sgdf}
S_{DIS} =  \ln\big(x\, g(x,\mu) \big).
\end{align}
This is in agreement with the result mentioned in the introduction, namely, that also the DIS entropy is not a directly observable quantity.

Based on AdS/CFT duality concepts, the correspondence between a gravity theory in a five-dimensional anti-de Sitter (AdS) space and conformal field theories (CFT) in physical space-time ~\cite{Maldacena:1997re, Gubser:1998bc, Witten:1998qj}, a semiclassical model has been developed~\cite{Brodsky:2006uqa,deTeramond:2008ht,physrep} which reproduces not only the hadron spectra from  hadronic two point functions but also allows one to obtain analytical expressions  for classical three point expectation values in 5-dimensional AdS metric. From these expressions follow three point functions  of hadronic states with vector or tensor currents  in the full momentum transfer range. From those  one can obtain form factors, Regge trajectories and particularly the intercept of the trajectory, $\alpha(0)$.  For the vector-isovector current the intercept is fixed by the spectra to $\alpha_\rho(0) = 0.5$ in fair agreement with phenomenology~\cite{deTeramond:2018ecg}. For the gravitational, {\it i.~e.} a spin-two current, Pomeron exchange is identified as the graviton of the dual AdS theory~\cite{deTeramond:2021lxc, Brower:2006ea}.

In the light front holographic approach the form factors can be expressed by universal functions of the generalized parton distributions at zero skewness~\cite{deTeramond:2018ecg} which yields, for the gluon density at small $x$, the relation~\cite{deTeramond:2021lxc}
\begin{align}  \lb{rel}
 x\,g(x,\mu) \sim \left(\frac{1}{x}\right)^{\alpha_P(0)-1},  \qquad x\ll1 ,
\end{align}
in the hadronic domain $\mu \approx 1$ GeV. This calculated density sets the initial scale for the DGLAP evolution equations. From~\req{Sgdf} and  \req{rel}  we then obtain
\begin{align} \lb{entro2}
S_{DIS} =  \ln \big( x\,  g(x,\mu) \big)
\sim   (\alpha_P(0)-1) \ln \Big(\frac{1}{ x} \Big),   \qquad x \ll 1.
\end{align}
It shows that a Pomeron intercept larger than 1 (hypercritical Pomeron), which leads to the rising of the total proton-proton cross section at large energies  (Fig.~\ref{regpp}), is uniquely related to a positive von Neumann entropy since the rapidity $Y =  \ln (1/ x)$ is always positive for $x < 1$. It is, in turn, a consequence of the separation of entangled parton states in a DIS experiment which gives rise to the entanglement entropy $S_{DIS}$.

The longitudinal light-front momentum fraction  $x$ can be identified with the Bjorken variable  $x_{bj}$ measured in the deep inelastic lepton-hadron scattering experiment
\begin{align} \lb{xbj}
x_{bj} =  \frac{Q^2}{2p \cdot q} =  \frac{Q^2}{W^2+Q^2-m_p^2},
\end{align} 
where $W^2 = (q + p)^2$ represents the total photon-hadron energy squared and $Q^2 = - q^2 > ~ 0$ is the photon virtuality. 
In the high-energy kinematic domain, $W^2 \gg Q^2$,  Eq.~\ref{xbj} reduces to $x = \frac{Q^2}{W^2}$ and, for fixed $Q^2$, we obtain
\begin{align} \lb{SDISW}
S_{DIS} \sim  (\alpha_P(0)-1)  \ln \left( \frac{W^2}{Q^2} \right) ,   \qquad W^2 \gg Q^2.
\end{align} 
It leads to the  remarkable result that one can identity  the value of $\alpha_P(0)$, which determines  the gluon distribution  $x g(x,\mu)$ at small $x$ in Eq.~\req{rel}, with the value of $\alpha_P(0)$, obtained from the high energy Pomeron contribution to the  $W$ dependence of the DIS cross section. 

The quark  distributions evaluated  in the light front holographic approach are in agreement with global fits~\cite{deTeramond:2018ecg, Liu:2019vsn} and have recently been extended to predict the strange and charm distributions in the proton~\cite{Sufian:2018cpj,Sufian:2020coz}. Likewise, the gluon distribution depends crucially on the relation \req{rel} and the value of the soft Pomeron intercept $\alpha_P(0)=1.086$. It has been inserted as the starting distribution  at the hadron scale for the DGLAP  evolution and also compares favorably with global fits without any additional ingredient~\cite{deTeramond:2021lxc}. This opens the way for the attractive possibility that there is only a single Pomeron  with a scale dependent intercept: It  manifests itself as a soft Pomeron at soft scales and hard Pomeron at harder ones, as was already considered in Refs.~\cite{DF15, pomhol}. In fact, as we discuss in the following subsection, the relation between the Pomeron intercept and the DIS entanglement entropy makes this possibility even more  compelling.    Also in the framework of the holographic approach we further discuss in Sec. II~D  the relation of hadron multiplicity with entanglement entropy first discussed in~\cite{kl1, kl2, KL, Hentschinski:2023izh}. In the final section III we point out also other relations between a hypercritical Pomeron intercept and rising entanglement entropy in high energy scattering in an asymptotically free theory.

\subsection{The scale dependent Pomeron \lb{sdP}}

Results from quasi-elastic photo and electroproduction of vector mesons  show,  for all processes at center of mass energies $W$ larger  than 10 GeV above threshold, an energy dependence of the integrated cross sections that can be fitted reasonably well by a power law $\sigma_{\gamma^*+p \to  \, V+p}  \approx W^\delta$. The value of $\delta $ depends strongly on the process and the virtuality of the photon as shown in Table \ref{xtab}. For integrated exclusive cross sections the whole trajectory contributes. In order to relate the exclusive cross sections, integrated over the whole range of momentum transfer $t$ to the Pomeron intercept  at $t=0$ one has to take into account  the   shrinkage corrections due to the slope of the trajectory. The latter can be estimated, and from this, one can reconstruct the energy dependence of the squared forward amplitude which, in Regge theory, is given by $\delta_{Regge} = 4 (\alpha_P(0)-1)$. The values for this quantity are  shown in Table~\ref{xtab}. 
\begin{table}[htp]
\begin{center}
\begin{tabular}{|c|c|c|c|}
\hline \hline
Reaction&$ \delta $& $\alpha_P-1$ & $ \chi^2$ \\
\hline
$\gamma \,p \to \rho\,p $& 0.190& 0.090 & 0.64 \\
$\gamma^* \,p \to \rho\,p  \, (Q^2 = 6 \, {\rm  GeV}^2)$ & 0.486& 0.135 & 0.66 \\
$\gamma \,p \to J/\psi\,p $& 0.675&  0.173 & 0.81\\
$\gamma \,p \to \Upsilon\,p $& 0.8626& 0.260 & 0.46 \\
\hline  \hline
\end{tabular}
\end{center}
\caption{\lb{xtab} Pomeron intercept from integrated cross sections for photo and electroproduction of vector mesons. The cross sections were fitted by the curve $\sigma = (W/W_0)^{\delta}$. The intercept $\alpha_P$ was obtained by taking into account the shrinkage corrections from the Regge trajectory slope, see \cite{DF15}.}
\end{table}

The situation is similar for the inclusive cross section for the DIS process $\gamma^*+p \to X$.  The proton structure function $F_2(x,Q^2)$ has been studied over a large range of values for $x$ (see~\req{xbj}) and $Q^2$ and the results for the $x$-dependence of the structure function could  be fitted by the power behavior~\cite{H1:2001ert}
\begin{align} \lb{lambH1}
F_2   \sim x g\left(x, Q^2\right) \sim x^{-\lambda(Q^2)}, ~~\mbox{with} ~~  \lambda(Q^2) = 0.048 \ln\left(\frac{Q^2}{\Lambda^2}\right),
\end{align}  
and $\Lambda \simeq 290$  MeV. The structure function is related to the total DIS cross section for the process $\gamma^* +p \to X$ by  
\begin{align} 
\sigma_{\gamma^*+p \to X}(W^2)  =  \frac{4 \pi^2 \alpha_{em}}{Q^2} F_2 \left(x, Q^2\right).
\end{align}
If one applies Regge theory to the elastic forward scattering amplitude $\gamma^*p \to \gamma^* p$ and relates it by the optical theorem to the DIS cross section $\gamma^* + p \to X$,  one expects a dependence  $\sigma_{\gamma^*+p \to X} \sim (W^2)^\lambda$, with $\lambda= \alpha_P(0)-1$, whereas the observed value of $\lambda$ varies between $\lambda = 0.16 \pm 0.015$ at $Q^2 = 2$ GeV$^2$  and  $\lambda = 0.36 \pm 0.11$  at $Q^2=150$ GeV$^2$~\cite{H1:1993jmo, ZEUS:1993ppj, H1:2001ert}.

This breakdown of Regge theory in electromagnetic processes led not to a crisis of this theory, since Regge theory was created for purely hadronic processes. But on the other hand, in quantum field theory a photon couples inevitably to intermediate hadronic matter and the lifetime of these intermediate states is large as compared to the time scale of a hard scattering process~\cite{Ioffe:1969kf}; In fact, in the case of vector meson production the lifetime of the hadronic state is even infinitely long. Therefore DIS scattering of (virtual) photons and diffractive electroproduction should be treated like hadron-hadron scattering and Regge theory should be applicable.

It was indeed shown by Donnachie and Landshoff that Regge theory can be applied to these processes by introducing two Pomerons \cite{Donnachie:1998gm}. One Pomeron was the usual ``soft’’ Pomeron, well established in hadron scattering, the other one was tentatively related to the so called BFKL \cite{Kuraev:1977fs, Balitsky:1978ic} ``hard” Pomeron~\cite{Ioffe:1969kf}. This emerges in perturbative QCD by the exchange of gluon ladders and leads to the effective power  $\lambda_h$
\begin{align} \lb{Del}
\lambda_{h} \equiv \alpha_{h} -1 = \frac{4 N_c \alpha_s}{\pi} \ln 2,
\end{align}
where $\alpha_h$ is the hard Pomeron intercept and $\alpha_s$ is the QCD coupling constant. This value is rather unstable against higher order contributions, but with resummation corrections an intercept between 1.3 and 1.5 is plausible.  With the conventional soft Pomeron and a BFKL Pomeron, with an intercept $\alpha^h\approx 1.42$, all  electron and photoproduction data in the available energy range could  be fitted with reasonable accuracy~\cite{Donnachie:2002en}. It turned out that the soft Pomeron couples dominantly  to extended objects and the hard one to smaller ones~\cite{Donnachie:2001wt}.

In DIS production the effective size of the intermediate hadronic system coupled to the virtual photon decreases with increasing virtuality. If one identifies the power $\lambda(Q^2)$  obtained from DIS experiments~\req{lambH1} with an effective scale-dependent Pomeron intercept  
\begin{align} \lb{laQ}
\lambda(Q^2) = \alpha_P(0,Q^2) -1,
\end{align}
this effective intercept increases indeed by a factor 2  from $Q^2 = 2$ GeV$^2$ to $Q^2 =150$ GeV$^2$. For the photoproduction of heavy vector mesons the transverse size of the vector meson is relevant, and is roughly proportional to the inverse mass and for a $J/\Psi$ and $\Upsilon$ meson, smaller than a normal hadronic size. As mentioned above, this effective intercept could for all energies available before LHC be explained by the two Pomeron model \cite{Donnachie:1998gm, Donnachie:2002en, Chwastowski:2003aw}. 

\begin{figure}[h]
\centering
\includegraphics[width=7.8cm]{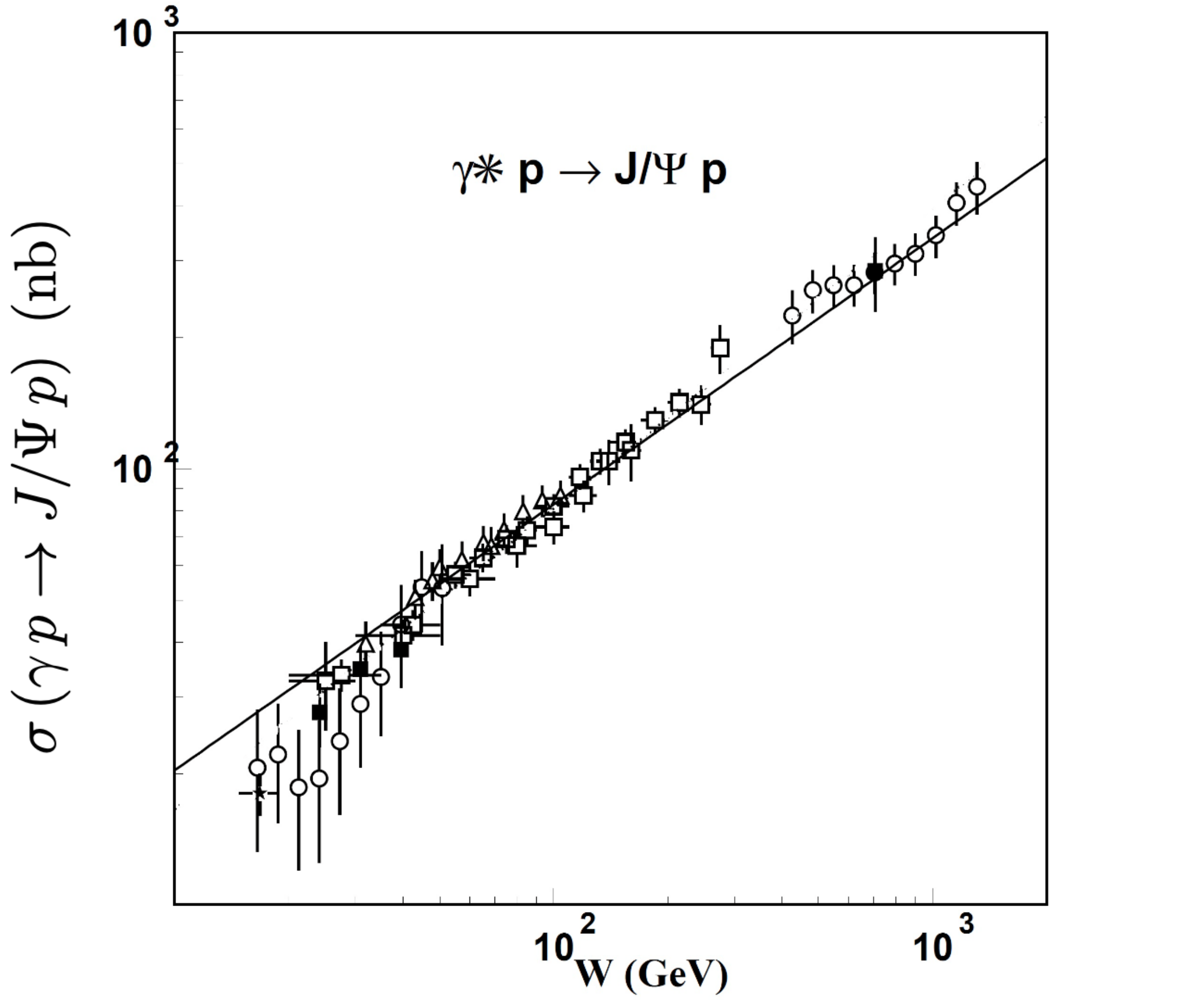} 
\caption{\lb{JPsiW} In photoproduction of hadrons the scale is set by the  virtuality of the mass of the produced $J/\psi$ meson, $Q^2 \simeq M^2_{J/\Psi}$. The production cross section for $J/\psi$ mesons can be described by a single Pomeron exchange  with intercept $\alpha_P(0) \simeq 1.15$ (solid line), distinctly higher than the intercept of conventional hadrons.  The  figure is adapted from \cite{DF15}, LHC data from~\cite{LHCb:2014acg, ALICE:2014eof}.}
\end{figure}

The two Pomeron picture, however, seems to be ruled out by the appearance of  data for quasi-elastic photoproduction of $J/\Psi$~\cite{LHCb:2014acg, ALICE:2014eof} and $\Upsilon$~\cite{LHCb:2015wlx} vector mesons at LHC energies. As can be seen from Fig.~\ref{JPsiW}, the power law for $J/\psi$ production holds over the range from 20 GeV to above 1 TeV. This is impossibly to achieve with two Pomerons, since at higher energies the hard one would dominate and the curve should show clear convexity. This leads to the introduction of a single, but scale dependent Pomeron~\cite{DF15} on a purely phenomenological basis. We thus write the $Q$-dependent effective power of the proton structure function $F_2(x,Q^2) \propto x^{\lambda(Q^2)}$ with the scale dependent Pomeron intercept $\alpha_P(0,Q^2)$ given by \req{laQ}, and the experimentally determined function $\lambda(Q^2)$ \req{lambH1} from~\cite{H1:2001ert},  as shown in Figure~\ref{lamu}.

In Ref.~\cite{pomhol} we have examined the possibility  that Eq.~\req{rel} is not only valid at the hadronic scale, but at all scales, thus leading to a single scale dependent Pomeron, $\alpha_P(0, \mu)$. Since the small-$x$ behavior is determined uniquely by the scale-dependent intercept it can, in principle, be extracted unambiguously.

\begin{figure}[h]
\centering
\includegraphics[width=7.8cm]{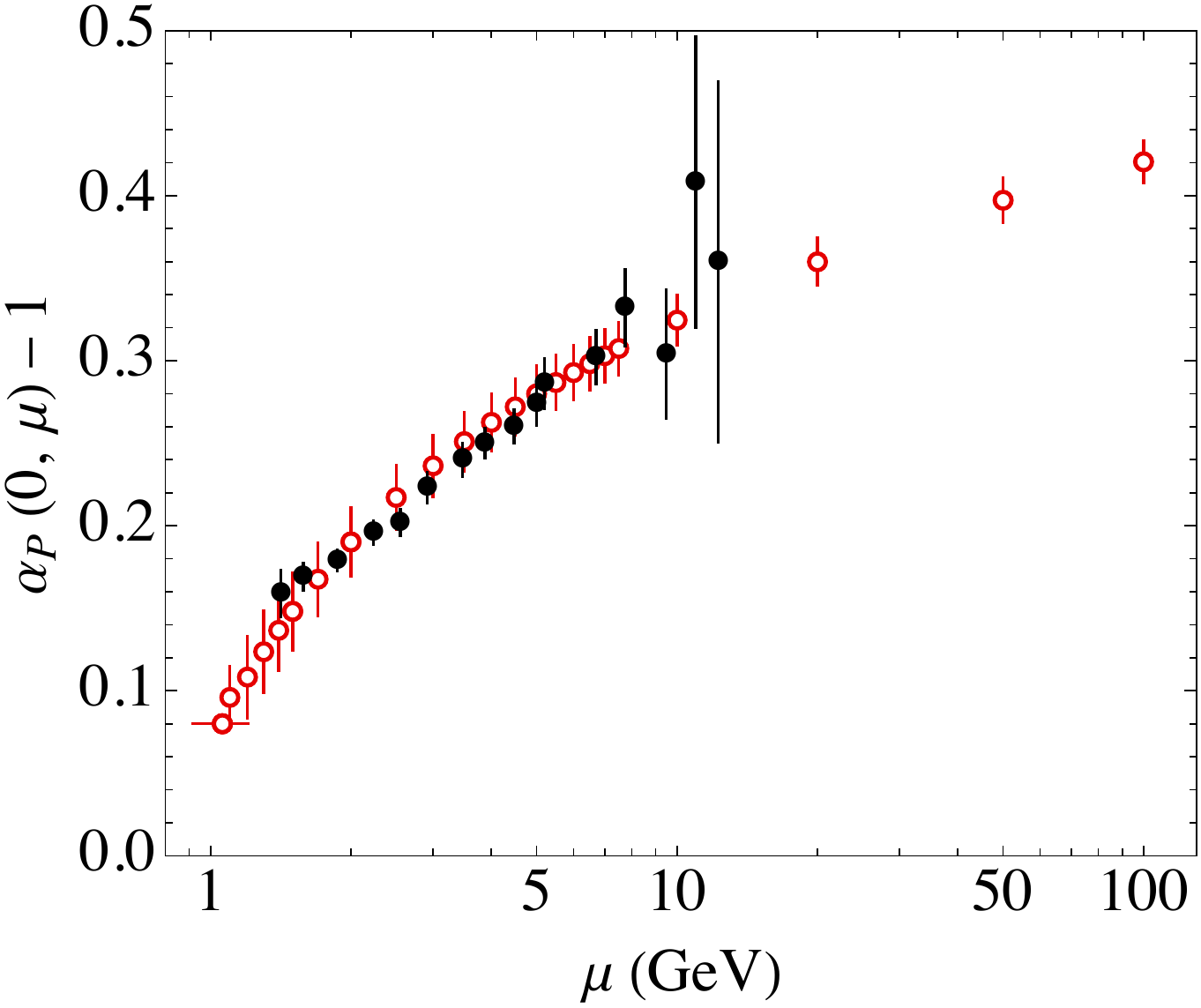}
\caption{\lb{lamu}  The effective power  $\lambda(\mu) = \alpha_P(0, \mu)- 1$ extracted from the  gluon distribution $g(x, \mu)$  for the pion (empty red  circles) using the procedure in~\cite{pomhol}, is compared with the values of $\lambda(Q^2)$ obtained from the measured proton structure function $F_2(x, Q^2)$  under the assumption $\mu^2=Q^2$. Experimental results from HERA~\cite{H1:2001ert} (filled black circles).}
\end{figure}

For very small values of $x$ the structure function $F_2(x,Q^2)$ is strongly dominated by the gluon distribution and therefore the $x$ dependence obtained in the  HERA analysis~\cite{H1:2001ert}, and parametrized in \req{lambH1}, yields a model independent determination of the Pomeron intercept if one accepts the relation between DIS-entropy, gluon density and Pomeron intercept expressed in \req{entro-c}. This allows the following comparison: The gluon distribution function obtained in 
\cite{deTeramond:2021lxc} can be evolved using the DGLAP equations and the scale dependence of the intercept can be obtained as described above. This is illustrated in Fig.~\ref{lamu}, where we compare  the values for our evolved results for the effective power  $\lambda(\mu)$, Eq.~\req{laQ},  in the range  $0.0001 \le x \le 0.00016$ following the procedure described in~\cite{pomhol},  with the power $\lambda(Q^2)$ of the proton structure function $F_2$ from Ref.~\cite{H1:2001ert}.  We show only the evolution results for the pion, since the proton results present numerical instabilities at lower evolution scales, currently under investigation. It should be noted, however, that this is rather a consistency check than an independent determination, since one has to relate the virtuality scale $\mu$ in the gluon distribution function   $g(x, \mu)$  to the photon virtuality $Q$ in the DIS process~\cite{H1:2001ert, pomhol} and the measured structure function is an important input for the evolution equations determining the $\mu$ dependence of the gluon distribution $g(x,\mu)$.

\subsection{\lb{multp} Hadron entropy and multiplicity}

It was suggested in Ref.~\cite{kl1}  a relation between the hadron multiplicity in the final state of the collision and the multiplicity of partons 
\begin{align} \lb{Sparhad}
S_{partons} = \ln\big(x g(x, Q^2) \big) \equiv S_{hadrons}.
\end{align}
Such a conjectural relation would be an indication of the absence of a significant increase of entropy in the hadronization process, as measured in the hadron multiplicity distributions~\cite{kl1, kl2, Gotsman:2020bjc}. In the present approach, this connection leads to the unexpected relevance of hadron multiplicity to the scale dependence of the Pomeron intercept,  and fixes an upper bound to the growth’s rate of hadron multiplicity at higher energies.

The theoretical analysis of hadron multiplicity distributions $P_n = \sigma_n/\sigma_{in}$,  the ratio of the cross section to produce $n$ hadrons in a high-energy collision to the inelastic cross section, is a rather complex undertaking~\cite{Koba:1972ng, Grosse-Oetringhaus:2009eis}. Most of the hadrons in the collision process are produced at relatively low transverse momentum and one has to recur to nonperturbative models or Monte Carlo event generator models of particle production.

A simple model which incorporates some of the main features of the particle multiplicity distributions is the parton cascade model of Levin and Lublinski~\cite{Levin:2003nc} for $P_n$:
\begin{align} \lb{dP}
\frac{d P_n(Y)}{dY} =  - \lambda_h n P_n(Y) + (n - 1) \lambda_h P_{n-1}(Y),
\end{align}
with the solution
\begin{align} \lb{Pn}
P_n(Y) =  e^{- \lambda_h Y} \left(1 - e^{- \lambda_h Y} \right)^{n - 1},
\end{align}
where  $Y = \ln(1/x)$ is the rapidity and $\lambda_h$ is the hard Pomeron intercept \req{Del}. The mean multiplicity which follows from \req{dP}  and \req{Pn} is
\begin{align}
\langle n \rangle  =  \sum_n n P_n(x) = e^{\lambda_h Y} = \left(\frac{1}{x}\right)^{\lambda_h},
\end{align}
which reproduces the power increase of the cross section with energy~\cite{kl1,Levin:2003nc}.\footnote{It was suggested in Ref.~\cite{Stoffers:2012mn} that the quantum entropy carried by the Pomeron to be at the origin of the entropy of the final state.}

In the present approach the gluon distribution at small $x$ is given by Eq.~\req{rel} and thus the multiplicity at the observational scale $\mu$ follows from Eq~\req{Sparhad} as discussed in \cite{kl1}
\begin{equation} \lb{nx}
x  g(x, \mu)  \sim \langle n \rangle = \sum_n n P_n = \left( \frac{1}{x} \right)^{\alpha_P(0, \mu) - 1},
\end{equation}
but with the Pomeron intercept $\lambda(\mu) = \alpha_P(0, \mu) - 1$ instead of $\lambda_h$. At small values of $x$ we have~\cite{Germano:2021brq, Bartels:2002uf, Carvalho:2007cf}  $x = \frac{t_0}{s}$, where we take the virtuality scale $\mu$ as the average transverse momentum $t_0$ of the produced hadrons or a particle mass in the final state of the collision~\cite{Germano:2021brq}.  Eq.  \req{nx} suggests that
\begin{align} \lb{ns}
\langle n \rangle  \sim  \left(\frac{s}{t_0}\right)^{\alpha_P(0,t_0)~-~1},
\end{align}
where  $t_0$  is independent of the collision energy $s$, but may differ for different datasets~\cite{Germano:2021brq}.
We notice that the theoretical predictions for the evolution of the effective power  $\lambda(\mu)$, shown in Fig.~\ref{lamu}, suggest that this quantity saturates at higher scales, setting an asymptotic upper bound for the rate of change of the hadron multiplicity \req{ns} for $\lambda \simeq 0.4 - 0.5$, higher than the value $\lambda = 1/3$ assumed in \cite{kl1} from  two-dimensional conformal field theory.

\begin{figure}[h]
\centering
\includegraphics[width=7.8cm]{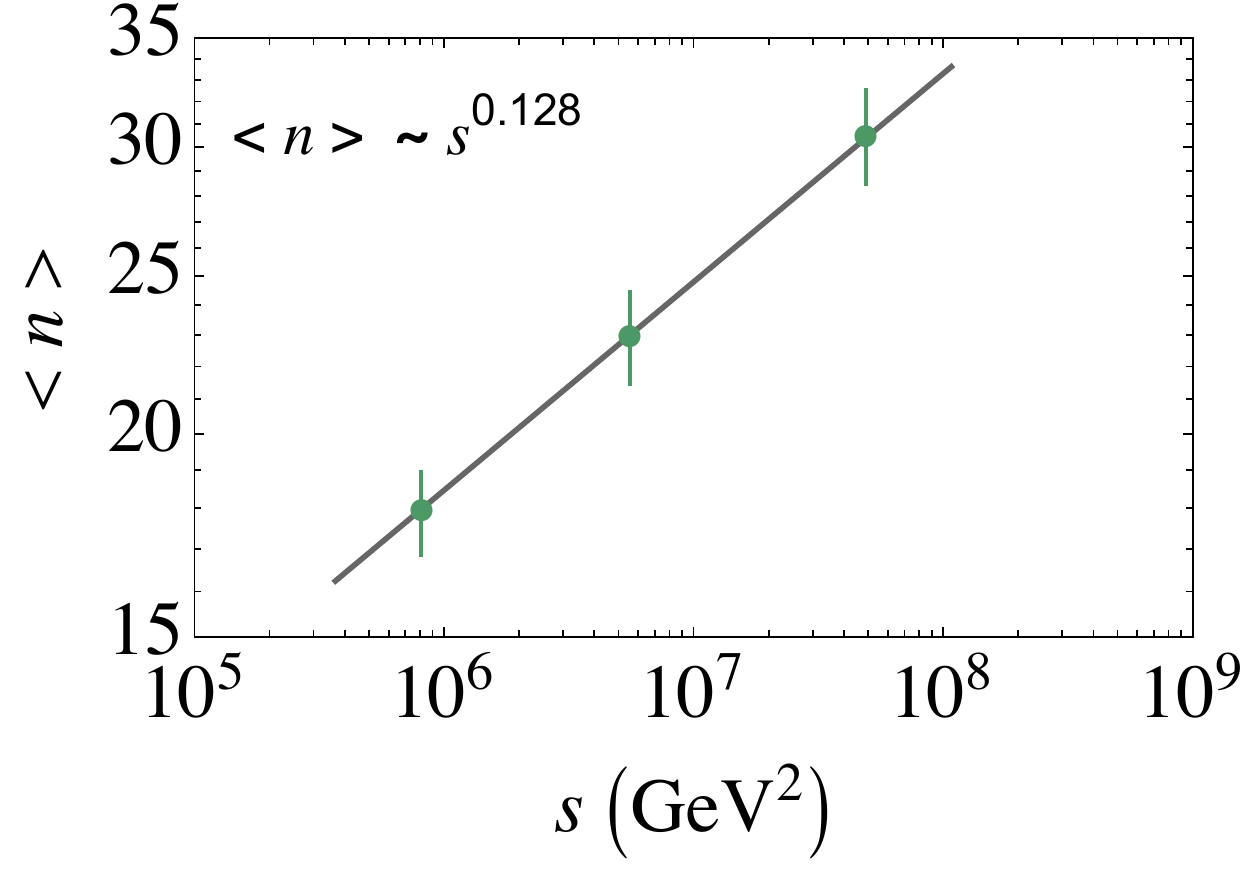} 
\includegraphics[width=7.8cm]{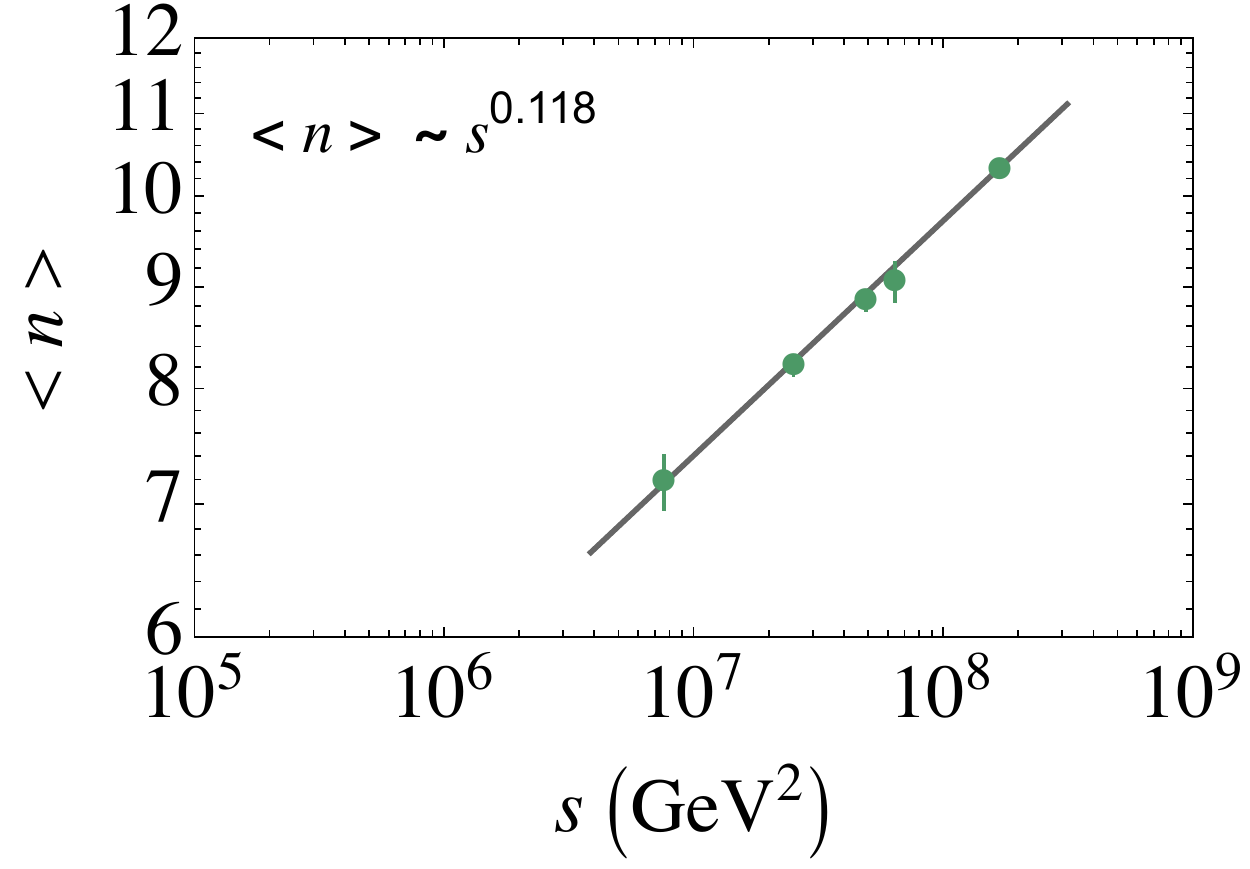} 
\caption{\lb{mltpfig}  Left figure: CMS data with $p_T > 100 $ MeV and pseudorapidity  $| \eta| < 0.5$  from~\cite{CMS:2010qvf}. Right  figure:  ALICE data with $p_T > 150$ MeV and $|\eta| < 0.8$ from~\cite{ALICE:2022xip}. The  grey lines in the figure correspond to Eq.~\req{ns} with $\sqrt{t_0} \simeq 1.3 - 1.4$ GeV.}
\end{figure}

In a recent analysis of LHC data~\cite{Germano:2021brq} it was found that the average multiplicity is well described by a power law, but with different powers for each dataset. As an illustration, we show in Fig.~\ref{mltpfig} the CMS  results from Ref.~\cite{CMS:2010qvf} which correspond to the Set I examined in Ref.~\cite{Germano:2021brq} for $p_T > 100$ MeV and pseudorapidity  $|\eta| < 0.5$.  The particles from this data set correspond to  gluon production. Set II from ALICE and ATLAS with larger rapidities $|\eta| < 2.4$ also includes the current fragmentation region, and Set III from ATLAS and CMS,  also with large rapidities, contains also hadrons produced in the perturbative domain. We thus examine here only the results for Set I from Ref.~\cite{Germano:2021brq} which abides to the mechanisms and approximations discussed in this article.  We also show in Fig.~\ref{mltpfig} more recent results from ALICE~\cite{ALICE:2022xip} with $p_T > 150$ MeV and $|\eta| < 0.8$ not included in~\cite{Germano:2021brq}.  Using the predictions from Eq.~\req{ns} (grey line in Fig.~\ref{mltpfig}), and comparing with the proton evolved results shown in Fig.~\ref{lamu}, we obtain for the average transverse momentum $\sqrt{t_0} \simeq 1.3 -1.4$ GeV.

Results for the charm structure function~\cite{H1:1996naa, ZEUS:1999qnm} $F^{c \bar c}(x,Q^2)$ indicate that in the presence of heavy quarks, the effective power $\lambda(Q^2)$ for the $x$ dependence~\req{lambH1} is larger than for the full structure function $F_2(x,Q^2)$ (see \cite{Donnachie:2001wt}).  This leads to the suggestion from Eq.~\req{ns}, that also the multiplicity of hidden charm objects increases faster than the multiplicity of hadrons with light quarks since, in this case, the virtuality scale is determined by the charm mass.

\section{Summary and Conclusion}

The light-front holographic framework, with essential constraints imposed by the generalized Veneziano model~\cite{AG69,LP70}, successfully describes the gluon distribution $g(x, \mu)$ in nucleons and mesons at the hadronic scale $\mu = 1$ GeV.  An essential input is its relation of the small-$x$ behavior of the gluon distribution  to the Pomeron intercept $\alpha_P(0)$ from high-energy scattering~\cite{deTeramond:2021lxc}.  

The gluon distribution is a scale dependent quantity and the intrinsic distribution evaluated at the hadronic scale can be extended to other scales using  QCD evolution equations. In Ref.~\cite{pomhol} we have ventured that also the Pomeron intercept follows this evolution and is therefore scale dependent. Such a scale dependent Pomeron was proposed on purely phenomenological reasons in Ref.~\cite{DF15}, where it was shown that such a dependence is suggested by the results for the photoproduction of $J/\psi$ mesons and not in contradiction with the  principles of Regge theory. A scale dependent Pomeron trajectory is also a salient feature of the gauge-gravity correspondence~\cite{Brower:2006ea}.  At very small values of $x$ the gluon dominance leads to universal behavior, well captured by the present description, which allows us to relate different experimental results in the high energy domain.

The relation (\ref{entro2})  makes it clear that two features of high energy scattering, the hypercritical Pomeron intercept $\alpha_P(0) >1$ and the single but scale dependent Pomeron, here proposed,  are closely related to typical features of a non-Abelian theory like QCD. In fact, in analogy of the thermodynamical definition of classical entropy for a system in equilibrium
\begin{align}
dS = \frac{dQ_{therm}}{T},
\end{align}
 the relation (\ref{entro2}) can be written in the differential form
\begin{align} \label{entro-c}
d S_{DIS}= d \ln\left(x\,g(x,\mu)\right)) \sim 2\, \left(\alpha_P(0) - 1\right) \frac{dW}{W}.
\end{align}
It shows that the hypercritical intercept $\alpha_P(0)$ is related to the increase of the DIS-entropy and the gluon density $x\,g(x,\mu)$ with increasing energy  $W$, corresponding to decreasing  $x$. The other salient feature discussed here, the scale dependence of the intercept, is a consequence  of the scale dependence of the gluon distribution, which in turn is the consequence of the renormalization group of an UV stable renormalizable theory like QCD. It is therefore quite natural that in a nonperturbative holographic theory corresponding to a non-Abelian gauge theory \cite{physrep} these features emerge in a transparent way.

If the quantum  entropy $S_{DIS}$ of partons inside a hadron  equals the classical  hadron entropy  $S_{hadron}$
of the final state, as proposed in Ref.~\cite{kl1}, then the final state entropy is also scale dependent and, with it, also the hadron multiplicity which depends on the average momentum of the hadrons in the final state; here the transverse momentum could set the scale as discussed in Sec. \ref{multp}. It describes the rate of change of the multiplicity  in terms of the scale-dependent Pomeron, and it sets an upper bound to its increase from the saturation of the effective power $\lambda(\mu)$ at very large virtualities predicted from the QCD evolution of the gluon distributions in holographic QCD.

We conclude this letter with some speculative remarks on a possible analogy between the black-hole (or Bekenstein-Hawking) entropy\footnote{For a short non-technical review see Jacob D. Bekenstein, Bekenstein-Hawking entropy,~\href{http://www.scholarpedia.org/article/Bekenstein-Hawking_entropy}{Scholarpedia, 3(10):7375 (2008)}.} and  $S_{DIS}$, the quantum entropy from deep inelastic scattering.

The existence of  black-hole entropy is demanded by the laws of thermodynamics~\cite{Bekenstein:1972tm}, but the observation of microstates inside the black hole  seems to be  prevented by the horizon (no hair theorems). There is, however, a way out:  Using the temperature of the black hole, which can be determined from the Hawking radiation~\cite{Hawking:1975vcx} emitted  from it, the Bekenstein-Hawkings entropy can be related to the surface area $A$ of the black hole
\begin{align}
S_{BH} = \frac{A}{4 L_P^2},
\end{align}
where $L_P$ is the Planck length $L_P=\sqrt{G_N \hbar/ c^3}$.

On the other hand, the DIS entropy $S_{DIS}$ is generated by the separation of the hadron state into parton states by the deep inelastic process. It is, however, not a directly observable quantity, since the partons cannot escape the hadron due to confinement. It also has the consequence that the partons at a fixed value of the Bjorken variable $x$  are not directly countable but the number depends on the renormalization scale $\mu^2$.  

In our approach to the gluon density~\cite{deTeramond:2021lxc} the parton number at small $x$ is given by the Pomeron intercept, see~\req{rel}.
 At first sight this seems like a contradiction, since the gluon intercept, at least for energies in and below the TeV region ({\it i.e.,} the LHC range and below), can be observed experimentally, {\it e.g.,} from forward scattering experiments. The solution to this problem consists in the introduction of a scale dependence of the Pomeron intercept, $\alpha_P(t=0,\mu^2)$, as discussed extensively in Sec.~\ref{sdP} and proposed earlier on phenomenological grounds in Refs.~\cite{pomhol, DF15} .

In both cases, $S_{BH}$ and $S_{DIS}$, the determination of the entropy is based on theoretical arguments. A direct counting of the microstates is forbidden: By the event horizon in black holes and by confinement in DIS. Additional theoretical arguments, however, allow a relation to observable quantities, by Hawking radiation and high energy scattering.  It may give further support to the idea of a possible relation between confinement and black holes as may be realized in a holographic context~\cite{Witten:2001}.

\acknowledgments

We thank Tianbo Liu for providing us the DGLAP results for the light-front holographic model shown in Fig.~\ref{lamu}.    We also thank Ulli Marquard for drawing our attention to the possible parallel  of black-hole entropy.  SJB is supported in part by the Department of Energy Contract No. DE-AC02-76SF00515.

\bibliography{DISentropy}

\begin{thebibliography}{57}%
\makeatletter
\providecommand \@ifxundefined [1]{%
 \@ifx{#1\undefined}
}%
\providecommand \@ifnum [1]{%
 \ifnum #1\expandafter \@firstoftwo
 \else \expandafter \@secondoftwo
 \fi
}%
\providecommand \@ifx [1]{%
 \ifx #1\expandafter \@firstoftwo
 \else \expandafter \@secondoftwo
 \fi
}%
\providecommand \natexlab [1]{#1}%
\providecommand \enquote  [1]{``#1''}%
\providecommand \bibnamefont  [1]{#1}%
\providecommand \bibfnamefont [1]{#1}%
\providecommand \citenamefont [1]{#1}%
\providecommand \href@noop [0]{\@secondoftwo}%
\providecommand \href [0]{\begingroup \@sanitize@url \@href}%
\providecommand \@href[1]{\@@startlink{#1}\@@href}%
\providecommand \@@href[1]{\endgroup#1\@@endlink}%
\providecommand \@sanitize@url [0]{\catcode `\\12\catcode `\$12\catcode
  `\&12\catcode `\#12\catcode `\^12\catcode `\_12\catcode `\%12\relax}%
\providecommand \@@startlink[1]{}%
\providecommand \@@endlink[0]{}%
\providecommand \url  [0]{\begingroup\@sanitize@url \@url }%
\providecommand \@url [1]{\endgroup\@href {#1}{\urlprefix }}%
\providecommand \urlprefix  [0]{URL }%
\providecommand \Eprint [0]{\href }%
\providecommand \doibase [0]{https://doi.org/}%
\providecommand \selectlanguage [0]{\@gobble}%
\providecommand \bibinfo  [0]{\@secondoftwo}%
\providecommand \bibfield  [0]{\@secondoftwo}%
\providecommand \translation [1]{[#1]}%
\providecommand \BibitemOpen [0]{}%
\providecommand \bibitemStop [0]{}%
\providecommand \bibitemNoStop [0]{.\EOS\space}%
\providecommand \EOS [0]{\spacefactor3000\relax}%
\providecommand \BibitemShut  [1]{\csname bibitem#1\endcsname}%
\let\auto@bib@innerbib\@empty
\bibitem [{\citenamefont {Kharzeev}\ and\ \citenamefont {Levin}(2017)}]{kl1}%
  \BibitemOpen
  \bibfield  {author} {\bibinfo {author} {\bibfnamefont {D.~E.}\ \bibnamefont
  {Kharzeev}}\ and\ \bibinfo {author} {\bibfnamefont {E.~M.}\ \bibnamefont
  {Levin}},\ }\bibfield  {title} {\bibinfo {title} {{Deep inelastic scattering
  as a probe of entanglement}},\ }\href
  {https://doi.org/10.1103/PhysRevD.95.114008} {\bibfield  {journal} {\bibinfo
  {journal} {Phys. Rev. D}\ }\textbf {\bibinfo {volume} {95}},\ \bibinfo
  {pages} {114008} (\bibinfo {year} {2017})},\ \Eprint
  {https://arxiv.org/abs/1702.03489} {arXiv:1702.03489 [hep-ph]} \BibitemShut
  {NoStop}%
\bibitem [{\citenamefont {Kharzeev}\ and\ \citenamefont {Levin}(2021)}]{kl2}%
  \BibitemOpen
  \bibfield  {author} {\bibinfo {author} {\bibfnamefont {D.~E.}\ \bibnamefont
  {Kharzeev}}\ and\ \bibinfo {author} {\bibfnamefont {E.}~\bibnamefont
  {Levin}},\ }\bibfield  {title} {\bibinfo {title} {{Deep inelastic scattering
  as a probe of entanglement: Confronting experimental data}},\ }\href
  {https://doi.org/10.1103/PhysRevD.104.L031503} {\bibfield  {journal}
  {\bibinfo  {journal} {Phys. Rev. D}\ }\textbf {\bibinfo {volume} {104}},\
  \bibinfo {pages} {L031503} (\bibinfo {year} {2021})},\ \Eprint
  {https://arxiv.org/abs/2102.09773} {arXiv:2102.09773 [hep-ph]} \BibitemShut
  {NoStop}%
\bibitem [{\citenamefont {de~T\'eramond}\ \emph {et~al.}(2021)\citenamefont
  {de~T\'eramond}, \citenamefont {Dosch}, \citenamefont {Liu}, \citenamefont
  {Sufian}, \citenamefont {Brodsky},\ and\ \citenamefont
  {Deur}}]{deTeramond:2021lxc}%
  \BibitemOpen
  \bibfield  {author} {\bibinfo {author} {\bibfnamefont {G.~F.}\ \bibnamefont
  {de~T\'eramond}}, \bibinfo {author} {\bibfnamefont {H.~G.}\ \bibnamefont
  {Dosch}}, \bibinfo {author} {\bibfnamefont {T.}~\bibnamefont {Liu}}, \bibinfo
  {author} {\bibfnamefont {R.~S.}\ \bibnamefont {Sufian}}, \bibinfo {author}
  {\bibfnamefont {S.~J.}\ \bibnamefont {Brodsky}},\ and\ \bibinfo {author}
  {\bibfnamefont {A.}~\bibnamefont {Deur}} (\bibinfo {collaboration} {HLFHS}),\
  }\bibfield  {title} {\bibinfo {title} {{Gluon matter distribution in the
  proton and pion from extended holographic light-front QCD}},\ }\href
  {https://doi.org/10.1103/PhysRevD.104.114005} {\bibfield  {journal} {\bibinfo
   {journal} {Phys. Rev. D}\ }\textbf {\bibinfo {volume} {104}},\ \bibinfo
  {pages} {114005} (\bibinfo {year} {2021})},\ \Eprint
  {https://arxiv.org/abs/2107.01231} {arXiv:2107.01231 [hep-ph]} \BibitemShut
  {NoStop}%
\bibitem [{\citenamefont {Brodsky}\ and\ \citenamefont
  {de~T\'eramond}(2006)}]{Brodsky:2006uqa}%
  \BibitemOpen
  \bibfield  {author} {\bibinfo {author} {\bibfnamefont {S.~J.}\ \bibnamefont
  {Brodsky}}\ and\ \bibinfo {author} {\bibfnamefont {G.~F.}\ \bibnamefont
  {de~T\'eramond}},\ }\bibfield  {title} {\bibinfo {title} {{Hadronic spectra
  and light-front wave functions in holographic QCD}},\ }\href
  {https://doi.org/10.1103/PhysRevLett.96.201601} {\bibfield  {journal}
  {\bibinfo  {journal} {Phys. Rev. Lett.}\ }\textbf {\bibinfo {volume} {96}},\
  \bibinfo {pages} {201601} (\bibinfo {year} {2006})},\ \Eprint
  {https://arxiv.org/abs/hep-ph/0602252} {arXiv:hep-ph/0602252} \BibitemShut
  {NoStop}%
\bibitem [{\citenamefont {de~T\'eramond}\ and\ \citenamefont
  {Brodsky}(2009)}]{deTeramond:2008ht}%
  \BibitemOpen
  \bibfield  {author} {\bibinfo {author} {\bibfnamefont {G.~F.}\ \bibnamefont
  {de~T\'eramond}}\ and\ \bibinfo {author} {\bibfnamefont {S.~J.}\ \bibnamefont
  {Brodsky}},\ }\bibfield  {title} {\bibinfo {title} {{Light-front holography:
  A first approximation to QCD}},\ }\href
  {https://doi.org/10.1103/PhysRevLett.102.081601} {\bibfield  {journal}
  {\bibinfo  {journal} {Phys. Rev. Lett.}\ }\textbf {\bibinfo {volume} {102}},\
  \bibinfo {pages} {081601} (\bibinfo {year} {2009})},\ \Eprint
  {https://arxiv.org/abs/0809.4899} {arXiv:0809.4899 [hep-ph]} \BibitemShut
  {NoStop}%
\bibitem [{\citenamefont {Brodsky}\ \emph {et~al.}(2015)\citenamefont
  {Brodsky}, \citenamefont {de~T\'eramond}, \citenamefont {Dosch},\ and\
  \citenamefont {Erlich}}]{physrep}%
  \BibitemOpen
  \bibfield  {author} {\bibinfo {author} {\bibfnamefont {S.~J.}\ \bibnamefont
  {Brodsky}}, \bibinfo {author} {\bibfnamefont {G.~F.}\ \bibnamefont
  {de~T\'eramond}}, \bibinfo {author} {\bibfnamefont {H.~G.}\ \bibnamefont
  {Dosch}},\ and\ \bibinfo {author} {\bibfnamefont {J.}~\bibnamefont
  {Erlich}},\ }\bibfield  {title} {\bibinfo {title} {{Light-front holographic
  QCD and emerging confinement}},\ }\href
  {https://doi.org/10.1016/j.physrep.2015.05.001} {\bibfield  {journal}
  {\bibinfo  {journal} {Phys. Rept.}\ }\textbf {\bibinfo {volume} {584}},\
  \bibinfo {pages} {1} (\bibinfo {year} {2015})},\ \Eprint
  {https://arxiv.org/abs/1407.8131} {arXiv:1407.8131 [hep-ph]} \BibitemShut
  {NoStop}%
\bibitem [{\citenamefont {Dosch}\ \emph {et~al.}(2022)\citenamefont {Dosch},
  \citenamefont {de~T\'eramond}, \citenamefont {Liu}, \citenamefont {Sufian},
  \citenamefont {Brodsky},\ and\ \citenamefont {Deur}}]{pomhol}%
  \BibitemOpen
  \bibfield  {author} {\bibinfo {author} {\bibfnamefont {H.~G.}\ \bibnamefont
  {Dosch}}, \bibinfo {author} {\bibfnamefont {G.~F.}\ \bibnamefont
  {de~T\'eramond}}, \bibinfo {author} {\bibfnamefont {T.}~\bibnamefont {Liu}},
  \bibinfo {author} {\bibfnamefont {R.~S.}\ \bibnamefont {Sufian}}, \bibinfo
  {author} {\bibfnamefont {S.~J.}\ \bibnamefont {Brodsky}},\ and\ \bibinfo
  {author} {\bibfnamefont {A.}~\bibnamefont {Deur}} (\bibinfo {collaboration}
  {HLFHS}),\ }\bibfield  {title} {\bibinfo {title} {{Towards a single
  scale-dependent Pomeron in holographic light-front QCD}},\ }\href
  {https://doi.org/10.1103/PhysRevD.105.034029} {\bibfield  {journal} {\bibinfo
   {journal} {Phys. Rev. D}\ }\textbf {\bibinfo {volume} {105}},\ \bibinfo
  {pages} {034029} (\bibinfo {year} {2022})},\ \Eprint
  {https://arxiv.org/abs/2201.09813} {arXiv:2201.09813 [hep-ph]} \BibitemShut
  {NoStop}%
\bibitem [{\citenamefont {Tu}\ \emph {et~al.}(2020)\citenamefont {Tu},
  \citenamefont {Kharzeev},\ and\ \citenamefont {Ullrich}}]{KL}%
  \BibitemOpen
  \bibfield  {author} {\bibinfo {author} {\bibfnamefont {Z.}~\bibnamefont
  {Tu}}, \bibinfo {author} {\bibfnamefont {D.~E.}\ \bibnamefont {Kharzeev}},\
  and\ \bibinfo {author} {\bibfnamefont {T.}~\bibnamefont {Ullrich}},\
  }\bibfield  {title} {\bibinfo {title} {{Einstein-Podolsky-Rosen paradox and
  quantum entanglement at subnucleonic scales}},\ }\href
  {https://doi.org/10.1103/PhysRevLett.124.062001} {\bibfield  {journal}
  {\bibinfo  {journal} {Phys. Rev. Lett.}\ }\textbf {\bibinfo {volume} {124}},\
  \bibinfo {pages} {062001} (\bibinfo {year} {2020})},\ \Eprint
  {https://arxiv.org/abs/1904.11974} {arXiv:1904.11974 [hep-ph]} \BibitemShut
  {NoStop}%
\bibitem [{\citenamefont {Hentschinski}\ \emph {et~al.}(2023)\citenamefont
  {Hentschinski}, \citenamefont {Kharzeev}, \citenamefont {Kutak},\ and\
  \citenamefont {Tu}}]{Hentschinski:2023izh}%
  \BibitemOpen
  \bibfield  {author} {\bibinfo {author} {\bibfnamefont {M.}~\bibnamefont
  {Hentschinski}}, \bibinfo {author} {\bibfnamefont {D.~E.}\ \bibnamefont
  {Kharzeev}}, \bibinfo {author} {\bibfnamefont {K.}~\bibnamefont {Kutak}},\
  and\ \bibinfo {author} {\bibfnamefont {Z.}~\bibnamefont {Tu}},\ }\bibfield
  {title} {\bibinfo {title} {{Probing the onset of maximal entanglement inside
  the proton in diffractive deep inelastic scattering}},\ }\href
  {https://doi.org/10.1103/PhysRevLett.131.241901} {\bibfield  {journal}
  {\bibinfo  {journal} {Phys. Rev. Lett.}\ }\textbf {\bibinfo {volume} {131}},\
  \bibinfo {pages} {241901} (\bibinfo {year} {2023})},\ \Eprint
  {https://arxiv.org/abs/2305.03069} {arXiv:2305.03069 [hep-ph]} \BibitemShut
  {NoStop}%
\bibitem [{\citenamefont {Dosch}\ and\ \citenamefont {Ferreira}(2015)}]{DF15}%
  \BibitemOpen
  \bibfield  {author} {\bibinfo {author} {\bibfnamefont {H.~G.}\ \bibnamefont
  {Dosch}}\ and\ \bibinfo {author} {\bibfnamefont {E.}~\bibnamefont
  {Ferreira}},\ }\bibfield  {title} {\bibinfo {title} {{Diffractive
  electromagnetic processes from a Regge point of view}},\ }\href
  {https://doi.org/10.1103/PhysRevD.92.034002} {\bibfield  {journal} {\bibinfo
  {journal} {Phys. Rev. D}\ }\textbf {\bibinfo {volume} {92}},\ \bibinfo
  {pages} {034002} (\bibinfo {year} {2015})},\ \Eprint
  {https://arxiv.org/abs/1507.03442} {arXiv:1507.03442 [hep-ph]} \BibitemShut
  {NoStop}%
\bibitem [{\citenamefont {Regge}(1959)}]{Reg59}%
  \BibitemOpen
  \bibfield  {author} {\bibinfo {author} {\bibfnamefont {T.}~\bibnamefont
  {Regge}},\ }\bibfield  {title} {\bibinfo {title} {{Introduction to complex
  orbital momenta}},\ }\href {https://doi.org/10.1007/BF02728177} {\bibfield
  {journal} {\bibinfo  {journal} {Nuovo Cim.}\ }\textbf {\bibinfo {volume}
  {14}},\ \bibinfo {pages} {951} (\bibinfo {year} {1959})}\BibitemShut
  {NoStop}%
\bibitem [{\citenamefont {Chew}\ and\ \citenamefont {Frautschi}(1961)}]{CF61}%
  \BibitemOpen
  \bibfield  {author} {\bibinfo {author} {\bibfnamefont {G.~F.}\ \bibnamefont
  {Chew}}\ and\ \bibinfo {author} {\bibfnamefont {S.~C.}\ \bibnamefont
  {Frautschi}},\ }\bibfield  {title} {\bibinfo {title} {{Principle of
  equivalence for all strongly interacting particles within the $S$-matrix
  framework}},\ }\href {https://doi.org/10.1103/PhysRevLett.7.394} {\bibfield
  {journal} {\bibinfo  {journal} {Phys. Rev. Lett.}\ }\textbf {\bibinfo
  {volume} {7}},\ \bibinfo {pages} {394} (\bibinfo {year} {1961})}\BibitemShut
  {NoStop}%
\bibitem [{\citenamefont {Zyla}\ \emph {et~al.}(2020)\citenamefont {Zyla} \emph
  {et~al.}}]{PDG20}%
  \BibitemOpen
  \bibfield  {author} {\bibinfo {author} {\bibfnamefont {P.~A.}\ \bibnamefont
  {Zyla}} \emph {et~al.} (\bibinfo {collaboration} {Particle Data Group}),\
  }\bibfield  {title} {\bibinfo {title} {{Review of Particle Physics}},\ }\href
  {https://doi.org/10.1093/ptep/ptaa104} {\bibfield  {journal} {\bibinfo
  {journal} {PTEP}\ }\textbf {\bibinfo {volume} {2020}},\ \bibinfo {pages}
  {083C01} (\bibinfo {year} {2020})}\BibitemShut {NoStop}%
\bibitem [{\citenamefont {de~T\'eramond}\ \emph {et~al.}(2018)\citenamefont
  {de~T\'eramond}, \citenamefont {Liu}, \citenamefont {Sufian}, \citenamefont
  {Dosch}, \citenamefont {Brodsky},\ and\ \citenamefont
  {Deur}}]{deTeramond:2018ecg}%
  \BibitemOpen
  \bibfield  {author} {\bibinfo {author} {\bibfnamefont {G.~F.}\ \bibnamefont
  {de~T\'eramond}}, \bibinfo {author} {\bibfnamefont {T.}~\bibnamefont {Liu}},
  \bibinfo {author} {\bibfnamefont {R.~S.}\ \bibnamefont {Sufian}}, \bibinfo
  {author} {\bibfnamefont {H.~G.}\ \bibnamefont {Dosch}}, \bibinfo {author}
  {\bibfnamefont {S.~J.}\ \bibnamefont {Brodsky}},\ and\ \bibinfo {author}
  {\bibfnamefont {A.}~\bibnamefont {Deur}} (\bibinfo {collaboration} {HLFHS}),\
  }\bibfield  {title} {\bibinfo {title} {{Universality of generalized parton
  distributions in light-front holographic QCD}},\ }\href
  {https://doi.org/10.1103/PhysRevLett.120.182001} {\bibfield  {journal}
  {\bibinfo  {journal} {Phys. Rev. Lett.}\ }\textbf {\bibinfo {volume} {120}},\
  \bibinfo {pages} {182001} (\bibinfo {year} {2018})},\ \Eprint
  {https://arxiv.org/abs/1801.09154} {arXiv:1801.09154 [hep-ph]} \BibitemShut
  {NoStop}%
\bibitem [{\citenamefont {Froissart}(1961)}]{Fro61}%
  \BibitemOpen
  \bibfield  {author} {\bibinfo {author} {\bibfnamefont {M.}~\bibnamefont
  {Froissart}},\ }\bibfield  {title} {\bibinfo {title} {{Asymptotic behavior
  and subtractions in the Mandelstam representation}},\ }\href
  {https://doi.org/10.1103/PhysRev.123.1053} {\bibfield  {journal} {\bibinfo
  {journal} {Phys. Rev.}\ }\textbf {\bibinfo {volume} {123}},\ \bibinfo {pages}
  {1053} (\bibinfo {year} {1961})}\BibitemShut {NoStop}%
\bibitem [{\citenamefont {Martin}(1965)}]{Mar65}%
  \BibitemOpen
  \bibfield  {author} {\bibinfo {author} {\bibfnamefont {A.}~\bibnamefont
  {Martin}},\ }\bibfield  {title} {\bibinfo {title} {{Extension of the
  axiomatic analyticity domain of scattering amplitudes by unitarity-I.}},\
  }\href {https://doi.org/10.1007/BF02720568} {\bibfield  {journal} {\bibinfo
  {journal} {Nuovo Cim. A}\ }\textbf {\bibinfo {volume} {42}},\ \bibinfo
  {pages} {930} (\bibinfo {year} {1965})}\BibitemShut {NoStop}%
\bibitem [{\citenamefont {Heisenberg}(1952)}]{Hei52}%
  \BibitemOpen
  \bibfield  {author} {\bibinfo {author} {\bibfnamefont {W.}~\bibnamefont
  {Heisenberg}},\ }\bibfield  {title} {\bibinfo {title} {Mesonenerzeugung als
  stoßwellenproblem},\ }\href@noop {} {\bibfield  {journal} {\bibinfo
  {journal} {Z. Physik}\ }\textbf {\bibinfo {volume} {133}},\ \bibinfo {pages}
  {65} (\bibinfo {year} {1952})}\BibitemShut {NoStop}%
\bibitem [{\citenamefont {Altarelli}\ and\ \citenamefont
  {Parisi}(1977)}]{AP77}%
  \BibitemOpen
  \bibfield  {author} {\bibinfo {author} {\bibfnamefont {G.}~\bibnamefont
  {Altarelli}}\ and\ \bibinfo {author} {\bibfnamefont {G.}~\bibnamefont
  {Parisi}},\ }\bibfield  {title} {\bibinfo {title} {{Asymptotic freedom in
  parton language}},\ }\href {https://doi.org/10.1016/0550-3213(77)90384-4}
  {\bibfield  {journal} {\bibinfo  {journal} {Nucl. Phys. B}\ }\textbf
  {\bibinfo {volume} {126}},\ \bibinfo {pages} {298} (\bibinfo {year}
  {1977})}\BibitemShut {NoStop}%
\bibitem [{\citenamefont {Dokshitzer}(1977)}]{Dok77}%
  \BibitemOpen
  \bibfield  {author} {\bibinfo {author} {\bibfnamefont {Y.~L.}\ \bibnamefont
  {Dokshitzer}},\ }\bibfield  {title} {\bibinfo {title} {{Calculation of the
  structure functions for deep inelastic scattering and $e^+ e^-$ annihilation
  by perturbation theory in quantum chromodynamics}},\ }\href@noop {}
  {\bibfield  {journal} {\bibinfo  {journal} {Sov. Phys. JETP}\ }\textbf
  {\bibinfo {volume} {46}},\ \bibinfo {pages} {641} (\bibinfo {year}
  {1977})}\BibitemShut {NoStop}%
\bibitem [{\citenamefont {Abramowicz}\ \emph {et~al.}(2015)\citenamefont
  {Abramowicz} \emph {et~al.}}]{hera15}%
  \BibitemOpen
  \bibfield  {author} {\bibinfo {author} {\bibfnamefont {H.}~\bibnamefont
  {Abramowicz}} \emph {et~al.} (\bibinfo {collaboration} {H1, ZEUS}),\
  }\bibfield  {title} {\bibinfo {title} {{Combination of measurements of
  inclusive deep inelastic ${e^{\pm}p}$ scattering cross sections and QCD
  analysis of HERA data}},\ }\href
  {https://doi.org/10.1140/epjc/s10052-015-3710-4} {\bibfield  {journal}
  {\bibinfo  {journal} {Eur. Phys. J. C}\ }\textbf {\bibinfo {volume} {75}},\
  \bibinfo {pages} {580} (\bibinfo {year} {2015})},\ \Eprint
  {https://arxiv.org/abs/1506.06042} {arXiv:1506.06042 [hep-ex]} \BibitemShut
  {NoStop}%
\bibitem [{\citenamefont {Maldacena}(1998)}]{Maldacena:1997re}%
  \BibitemOpen
  \bibfield  {author} {\bibinfo {author} {\bibfnamefont {J.~M.}\ \bibnamefont
  {Maldacena}},\ }\bibfield  {title} {\bibinfo {title} {{The Large-N limit of
  superconformal field theories and supergravity}},\ }\href
  {https://doi.org/10.1023/A:1026654312961} {\bibfield  {journal} {\bibinfo
  {journal} {Adv. Theor. Math. Phys.}\ }\textbf {\bibinfo {volume} {2}},\
  \bibinfo {pages} {231} (\bibinfo {year} {1998})},\ \Eprint
  {https://arxiv.org/abs/hep-th/9711200} {arXiv:hep-th/9711200} \BibitemShut
  {NoStop}%
\bibitem [{\citenamefont {Gubser}\ \emph {et~al.}(1998)\citenamefont {Gubser},
  \citenamefont {Klebanov},\ and\ \citenamefont {Polyakov}}]{Gubser:1998bc}%
  \BibitemOpen
  \bibfield  {author} {\bibinfo {author} {\bibfnamefont {S.~S.}\ \bibnamefont
  {Gubser}}, \bibinfo {author} {\bibfnamefont {I.~R.}\ \bibnamefont
  {Klebanov}},\ and\ \bibinfo {author} {\bibfnamefont {A.~M.}\ \bibnamefont
  {Polyakov}},\ }\bibfield  {title} {\bibinfo {title} {{Gauge theory
  correlators from noncritical string theory}},\ }\href
  {https://doi.org/10.1016/S0370-2693(98)00377-3} {\bibfield  {journal}
  {\bibinfo  {journal} {Phys. Lett. B}\ }\textbf {\bibinfo {volume} {428}},\
  \bibinfo {pages} {105} (\bibinfo {year} {1998})},\ \Eprint
  {https://arxiv.org/abs/hep-th/9802109} {arXiv:hep-th/9802109} \BibitemShut
  {NoStop}%
\bibitem [{\citenamefont {Witten}(1998)}]{Witten:1998qj}%
  \BibitemOpen
  \bibfield  {author} {\bibinfo {author} {\bibfnamefont {E.}~\bibnamefont
  {Witten}},\ }\bibfield  {title} {\bibinfo {title} {{Anti-de Sitter space and
  holography}},\ }\href {https://doi.org/10.4310/ATMP.1998.v2.n2.a2} {\bibfield
   {journal} {\bibinfo  {journal} {Adv. Theor. Math. Phys.}\ }\textbf {\bibinfo
  {volume} {2}},\ \bibinfo {pages} {253} (\bibinfo {year} {1998})},\ \Eprint
  {https://arxiv.org/abs/hep-th/9802150} {arXiv:hep-th/9802150} \BibitemShut
  {NoStop}%
\bibitem [{\citenamefont {Brower}\ \emph {et~al.}(2007)\citenamefont {Brower},
  \citenamefont {Polchinski}, \citenamefont {Strassler},\ and\ \citenamefont
  {Tan}}]{Brower:2006ea}%
  \BibitemOpen
  \bibfield  {author} {\bibinfo {author} {\bibfnamefont {R.~C.}\ \bibnamefont
  {Brower}}, \bibinfo {author} {\bibfnamefont {J.}~\bibnamefont {Polchinski}},
  \bibinfo {author} {\bibfnamefont {M.~J.}\ \bibnamefont {Strassler}},\ and\
  \bibinfo {author} {\bibfnamefont {C.-I.}\ \bibnamefont {Tan}},\ }\bibfield
  {title} {\bibinfo {title} {{The Pomeron and gauge/string duality}},\ }\href
  {https://doi.org/10.1088/1126-6708/2007/12/005} {\bibfield  {journal}
  {\bibinfo  {journal} {J. High Energy Phys.}\ }\textbf {\bibinfo {volume}
  {12}},\ \bibinfo {pages} {005}},\ \Eprint
  {https://arxiv.org/abs/hep-th/0603115} {arXiv:hep-th/0603115} \BibitemShut
  {NoStop}%
\bibitem [{\citenamefont {Liu}\ \emph {et~al.}(2020)\citenamefont {Liu},
  \citenamefont {Sufian}, \citenamefont {de~T\'eramond}, \citenamefont {Dosch},
  \citenamefont {Brodsky},\ and\ \citenamefont {Deur}}]{Liu:2019vsn}%
  \BibitemOpen
  \bibfield  {author} {\bibinfo {author} {\bibfnamefont {T.}~\bibnamefont
  {Liu}}, \bibinfo {author} {\bibfnamefont {R.~S.}\ \bibnamefont {Sufian}},
  \bibinfo {author} {\bibfnamefont {G.~F.}\ \bibnamefont {de~T\'eramond}},
  \bibinfo {author} {\bibfnamefont {H.~G.}\ \bibnamefont {Dosch}}, \bibinfo
  {author} {\bibfnamefont {S.~J.}\ \bibnamefont {Brodsky}},\ and\ \bibinfo
  {author} {\bibfnamefont {A.}~\bibnamefont {Deur}} (\bibinfo {collaboration}
  {HLFHS}),\ }\bibfield  {title} {\bibinfo {title} {{Unified description of
  polarized and unpolarized quark distributions in the proton}},\ }\href
  {https://doi.org/10.1103/PhysRevLett.124.082003} {\bibfield  {journal}
  {\bibinfo  {journal} {Phys. Rev. Lett.}\ }\textbf {\bibinfo {volume} {124}},\
  \bibinfo {pages} {082003} (\bibinfo {year} {2020})},\ \Eprint
  {https://arxiv.org/abs/1909.13818} {arXiv:1909.13818 [hep-ph]} \BibitemShut
  {NoStop}%
\bibitem [{\citenamefont {Sufian}\ \emph {et~al.}(2018)\citenamefont {Sufian},
  \citenamefont {Liu}, \citenamefont {de~T\'eramond}, \citenamefont {Dosch},
  \citenamefont {Brodsky}, \citenamefont {Deur}, \citenamefont {Islam},\ and\
  \citenamefont {Ma}}]{Sufian:2018cpj}%
  \BibitemOpen
  \bibfield  {author} {\bibinfo {author} {\bibfnamefont {R.~S.}\ \bibnamefont
  {Sufian}}, \bibinfo {author} {\bibfnamefont {T.}~\bibnamefont {Liu}},
  \bibinfo {author} {\bibfnamefont {G.~F.}\ \bibnamefont {de~T\'eramond}},
  \bibinfo {author} {\bibfnamefont {H.~G.}\ \bibnamefont {Dosch}}, \bibinfo
  {author} {\bibfnamefont {S.~J.}\ \bibnamefont {Brodsky}}, \bibinfo {author}
  {\bibfnamefont {A.}~\bibnamefont {Deur}}, \bibinfo {author} {\bibfnamefont
  {M.~T.}\ \bibnamefont {Islam}},\ and\ \bibinfo {author} {\bibfnamefont
  {B.-Q.}\ \bibnamefont {Ma}},\ }\bibfield  {title} {\bibinfo {title}
  {{Nonperturbative strange-quark sea from lattice QCD, light-front holography,
  and meson-baryon fluctuation models}},\ }\href
  {https://doi.org/10.1103/PhysRevD.98.114004} {\bibfield  {journal} {\bibinfo
  {journal} {Phys. Rev. D}\ }\textbf {\bibinfo {volume} {98}},\ \bibinfo
  {pages} {114004} (\bibinfo {year} {2018})},\ \Eprint
  {https://arxiv.org/abs/1809.04975} {arXiv:1809.04975 [hep-ph]} \BibitemShut
  {NoStop}%
\bibitem [{\citenamefont {Sufian}\ \emph {et~al.}(2020)\citenamefont {Sufian},
  \citenamefont {Liu}, \citenamefont {Alexandru}, \citenamefont {Brodsky},
  \citenamefont {de~T\'eramond}, \citenamefont {Dosch}, \citenamefont {Draper},
  \citenamefont {Liu},\ and\ \citenamefont {Yang}}]{Sufian:2020coz}%
  \BibitemOpen
  \bibfield  {author} {\bibinfo {author} {\bibfnamefont {R.~S.}\ \bibnamefont
  {Sufian}}, \bibinfo {author} {\bibfnamefont {T.}~\bibnamefont {Liu}},
  \bibinfo {author} {\bibfnamefont {A.}~\bibnamefont {Alexandru}}, \bibinfo
  {author} {\bibfnamefont {S.~J.}\ \bibnamefont {Brodsky}}, \bibinfo {author}
  {\bibfnamefont {G.~F.}\ \bibnamefont {de~T\'eramond}}, \bibinfo {author}
  {\bibfnamefont {H.~G.}\ \bibnamefont {Dosch}}, \bibinfo {author}
  {\bibfnamefont {T.}~\bibnamefont {Draper}}, \bibinfo {author} {\bibfnamefont
  {K.-F.}\ \bibnamefont {Liu}},\ and\ \bibinfo {author} {\bibfnamefont {Y.-B.}\
  \bibnamefont {Yang}},\ }\bibfield  {title} {\bibinfo {title} {{Constraints on
  charm-anticharm asymmetry in the nucleon from lattice QCD}},\ }\href
  {https://doi.org/10.1016/j.physletb.2020.135633} {\bibfield  {journal}
  {\bibinfo  {journal} {Phys. Lett. B}\ }\textbf {\bibinfo {volume} {808}},\
  \bibinfo {pages} {135633} (\bibinfo {year} {2020})},\ \Eprint
  {https://arxiv.org/abs/2003.01078} {arXiv:2003.01078 [hep-lat]} \BibitemShut
  {NoStop}%
\bibitem [{\citenamefont {Adloff}\ \emph {et~al.}(2001)\citenamefont {Adloff}
  \emph {et~al.}}]{H1:2001ert}%
  \BibitemOpen
  \bibfield  {author} {\bibinfo {author} {\bibfnamefont {C.}~\bibnamefont
  {Adloff}} \emph {et~al.} (\bibinfo {collaboration} {H1}),\ }\bibfield
  {title} {\bibinfo {title} {{On the rise of the proton structure function
  $F_2$ towards low $x$}},\ }\href
  {https://doi.org/10.1016/S0370-2693(01)01074-7} {\bibfield  {journal}
  {\bibinfo  {journal} {Phys. Lett. B}\ }\textbf {\bibinfo {volume} {520}},\
  \bibinfo {pages} {183} (\bibinfo {year} {2001})},\ \Eprint
  {https://arxiv.org/abs/hep-ex/0108035} {arXiv:hep-ex/0108035} \BibitemShut
  {NoStop}%
\bibitem [{\citenamefont {Abt}\ \emph {et~al.}(1993)\citenamefont {Abt} \emph
  {et~al.}}]{H1:1993jmo}%
  \BibitemOpen
  \bibfield  {author} {\bibinfo {author} {\bibfnamefont {I.}~\bibnamefont
  {Abt}} \emph {et~al.} (\bibinfo {collaboration} {H1}),\ }\bibfield  {title}
  {\bibinfo {title} {{Measurement of the proton structure function $F_2(x,
  Q^2)$ in the low-$x$ region at HERA}},\ }\href
  {https://doi.org/10.1016/0550-3213(93)90090-C} {\bibfield  {journal}
  {\bibinfo  {journal} {Nucl. Phys. B}\ }\textbf {\bibinfo {volume} {407}},\
  \bibinfo {pages} {515} (\bibinfo {year} {1993})}\BibitemShut {NoStop}%
\bibitem [{\citenamefont {Derrick}\ \emph {et~al.}(1993)\citenamefont {Derrick}
  \emph {et~al.}}]{ZEUS:1993ppj}%
  \BibitemOpen
  \bibfield  {author} {\bibinfo {author} {\bibfnamefont {M.}~\bibnamefont
  {Derrick}} \emph {et~al.} (\bibinfo {collaboration} {ZEUS}),\ }\bibfield
  {title} {\bibinfo {title} {{Measurement of the proton structure function
  $F_2$ in $e p$ scattering at HERA}},\ }\href
  {https://doi.org/10.1016/0370-2693(93)90347-K} {\bibfield  {journal}
  {\bibinfo  {journal} {Phys. Lett. B}\ }\textbf {\bibinfo {volume} {316}},\
  \bibinfo {pages} {412} (\bibinfo {year} {1993})}\BibitemShut {NoStop}%
\bibitem [{\citenamefont {Ioffe}(1969)}]{Ioffe:1969kf}%
  \BibitemOpen
  \bibfield  {author} {\bibinfo {author} {\bibfnamefont {B.~L.}\ \bibnamefont
  {Ioffe}},\ }\bibfield  {title} {\bibinfo {title} {{Space-time picture of
  photon and neutrino scattering and electroproduction cross-section
  asymptotics}},\ }\href {https://doi.org/10.1016/0370-2693(69)90415-8}
  {\bibfield  {journal} {\bibinfo  {journal} {Phys. Lett. B}\ }\textbf
  {\bibinfo {volume} {30}},\ \bibinfo {pages} {123} (\bibinfo {year}
  {1969})}\BibitemShut {NoStop}%
\bibitem [{\citenamefont {Donnachie}\ and\ \citenamefont
  {Landshoff}(1998)}]{Donnachie:1998gm}%
  \BibitemOpen
  \bibfield  {author} {\bibinfo {author} {\bibfnamefont {A.}~\bibnamefont
  {Donnachie}}\ and\ \bibinfo {author} {\bibfnamefont {P.~V.}\ \bibnamefont
  {Landshoff}},\ }\bibfield  {title} {\bibinfo {title} {{Small $x$: Two
  pomerons!}},\ }\href {https://doi.org/10.1016/S0370-2693(98)00899-5}
  {\bibfield  {journal} {\bibinfo  {journal} {Phys. Lett. B}\ }\textbf
  {\bibinfo {volume} {437}},\ \bibinfo {pages} {408} (\bibinfo {year}
  {1998})},\ \Eprint {https://arxiv.org/abs/hep-ph/9806344}
  {arXiv:hep-ph/9806344} \BibitemShut {NoStop}%
\bibitem [{\citenamefont {Kuraev}\ \emph {et~al.}(1977)\citenamefont {Kuraev},
  \citenamefont {Lipatov},\ and\ \citenamefont {Fadin}}]{Kuraev:1977fs}%
  \BibitemOpen
  \bibfield  {author} {\bibinfo {author} {\bibfnamefont {E.~A.}\ \bibnamefont
  {Kuraev}}, \bibinfo {author} {\bibfnamefont {L.~N.}\ \bibnamefont
  {Lipatov}},\ and\ \bibinfo {author} {\bibfnamefont {V.~S.}\ \bibnamefont
  {Fadin}},\ }\bibfield  {title} {\bibinfo {title} {{The Pomeranchuk
  singularity in nonabelian gauge theories}},\ }\href@noop {} {\bibfield
  {journal} {\bibinfo  {journal} {Sov. Phys. JETP}\ }\textbf {\bibinfo {volume}
  {45}},\ \bibinfo {pages} {199} (\bibinfo {year} {1977})}\BibitemShut
  {NoStop}%
\bibitem [{\citenamefont {Balitsky}\ and\ \citenamefont
  {Lipatov}(1978)}]{Balitsky:1978ic}%
  \BibitemOpen
  \bibfield  {author} {\bibinfo {author} {\bibfnamefont {I.~I.}\ \bibnamefont
  {Balitsky}}\ and\ \bibinfo {author} {\bibfnamefont {L.~N.}\ \bibnamefont
  {Lipatov}},\ }\bibfield  {title} {\bibinfo {title} {{The Pomeranchuk
  singularity in quantum chromodynamics}},\ }\href@noop {} {\bibfield
  {journal} {\bibinfo  {journal} {Sov. J. Nucl. Phys.}\ }\textbf {\bibinfo
  {volume} {28}},\ \bibinfo {pages} {822} (\bibinfo {year} {1978})}\BibitemShut
  {NoStop}%
\bibitem [{\citenamefont {Donnachie}\ \emph {et~al.}(2004)\citenamefont
  {Donnachie}, \citenamefont {Dosch}, \citenamefont {Nachtmann},\ and\
  \citenamefont {Landshoff}}]{Donnachie:2002en}%
  \BibitemOpen
  \bibfield  {author} {\bibinfo {author} {\bibfnamefont {S.}~\bibnamefont
  {Donnachie}}, \bibinfo {author} {\bibfnamefont {H.~G.}\ \bibnamefont
  {Dosch}}, \bibinfo {author} {\bibfnamefont {O.}~\bibnamefont {Nachtmann}},\
  and\ \bibinfo {author} {\bibfnamefont {P.}~\bibnamefont {Landshoff}},\
  }\href@noop {} {\emph {\bibinfo {title} {{Pomeron physics and QCD}}}},\
  Vol.~\bibinfo {volume} {19}\ (\bibinfo  {publisher} {Cambridge University
  Press},\ \bibinfo {year} {2004})\BibitemShut {NoStop}%
\bibitem [{\citenamefont {Donnachie}\ and\ \citenamefont
  {Dosch}(2002)}]{Donnachie:2001wt}%
  \BibitemOpen
  \bibfield  {author} {\bibinfo {author} {\bibfnamefont {A.}~\bibnamefont
  {Donnachie}}\ and\ \bibinfo {author} {\bibfnamefont {H.~G.}\ \bibnamefont
  {Dosch}},\ }\bibfield  {title} {\bibinfo {title} {{Comprehensive approach to
  structure functions}},\ }\href {https://doi.org/10.1103/PhysRevD.65.014019}
  {\bibfield  {journal} {\bibinfo  {journal} {Phys. Rev. D}\ }\textbf {\bibinfo
  {volume} {65}},\ \bibinfo {pages} {014019} (\bibinfo {year} {2002})},\
  \Eprint {https://arxiv.org/abs/hep-ph/0106169} {arXiv:hep-ph/0106169}
  \BibitemShut {NoStop}%
\bibitem [{\citenamefont {Chwastowski}\ and\ \citenamefont
  {Figiel}(2004)}]{Chwastowski:2003aw}%
  \BibitemOpen
  \bibfield  {author} {\bibinfo {author} {\bibfnamefont {J.}~\bibnamefont
  {Chwastowski}}\ and\ \bibinfo {author} {\bibfnamefont {J.}~\bibnamefont
  {Figiel}},\ }\bibfield  {title} {\bibinfo {title} {{Photoproduction at
  HERA}},\ }\href@noop {} {\bibfield  {journal} {\bibinfo  {journal} {Phys.
  Part. Nucl.}\ }\textbf {\bibinfo {volume} {35}},\ \bibinfo {pages} {619}
  (\bibinfo {year} {2004})},\ \Eprint {https://arxiv.org/abs/hep-ex/0311044}
  {arXiv:hep-ex/0311044} \BibitemShut {NoStop}%
\bibitem [{\citenamefont {Aaij}\ \emph {et~al.}(2014)\citenamefont {Aaij} \emph
  {et~al.}}]{LHCb:2014acg}%
  \BibitemOpen
  \bibfield  {author} {\bibinfo {author} {\bibfnamefont {R.}~\bibnamefont
  {Aaij}} \emph {et~al.} (\bibinfo {collaboration} {LHCb}),\ }\bibfield
  {title} {\bibinfo {title} {{Updated measurements of exclusive $J/\Psi$ and
  $\Psi$(2S) production cross-sections in $pp$ collisions at $\sqrt{s}=7$
  TeV}},\ }\href {https://doi.org/10.1088/0954-3899/41/5/055002} {\bibfield
  {journal} {\bibinfo  {journal} {J. Phys. G}\ }\textbf {\bibinfo {volume}
  {41}},\ \bibinfo {pages} {055002} (\bibinfo {year} {2014})},\ \Eprint
  {https://arxiv.org/abs/1401.3288} {arXiv:1401.3288 [hep-ex]} \BibitemShut
  {NoStop}%
\bibitem [{\citenamefont {Abelev}\ \emph {et~al.}(2014)\citenamefont {Abelev}
  \emph {et~al.}}]{ALICE:2014eof}%
  \BibitemOpen
  \bibfield  {author} {\bibinfo {author} {\bibfnamefont {B.~B.}\ \bibnamefont
  {Abelev}} \emph {et~al.} (\bibinfo {collaboration} {ALICE}),\ }\bibfield
  {title} {\bibinfo {title} {{Exclusive $\mathrm{J/}\psi$ photoproduction off
  protons in ultraperipheral $p$-Pb collisions at $\sqrt{s_{\rm NN}}=5.02$
  TeV}},\ }\href {https://doi.org/10.1103/PhysRevLett.113.232504} {\bibfield
  {journal} {\bibinfo  {journal} {Phys. Rev. Lett.}\ }\textbf {\bibinfo
  {volume} {113}},\ \bibinfo {pages} {232504} (\bibinfo {year} {2014})},\
  \Eprint {https://arxiv.org/abs/1406.7819} {arXiv:1406.7819 [nucl-ex]}
  \BibitemShut {NoStop}%
\bibitem [{\citenamefont {Aaij}\ \emph {et~al.}(2015)\citenamefont {Aaij} \emph
  {et~al.}}]{LHCb:2015wlx}%
  \BibitemOpen
  \bibfield  {author} {\bibinfo {author} {\bibfnamefont {R.}~\bibnamefont
  {Aaij}} \emph {et~al.} (\bibinfo {collaboration} {LHCb}),\ }\bibfield
  {title} {\bibinfo {title} {{Measurement of the exclusive
  \ensuremath{\Upsilon} production cross-section in pp collisions at $
  \sqrt{s}=7 $ TeV and 8 TeV}},\ }\href
  {https://doi.org/10.1007/JHEP09(2015)084} {\bibfield  {journal} {\bibinfo
  {journal} {J. High Energy Phys.}\ }\textbf {\bibinfo {volume} {09}},\
  \bibinfo {pages} {084}},\ \Eprint {https://arxiv.org/abs/1505.08139}
  {arXiv:1505.08139 [hep-ex]} \BibitemShut {NoStop}%
\bibitem [{\citenamefont {Gotsman}\ and\ \citenamefont
  {Levin}(2020)}]{Gotsman:2020bjc}%
  \BibitemOpen
  \bibfield  {author} {\bibinfo {author} {\bibfnamefont {E.}~\bibnamefont
  {Gotsman}}\ and\ \bibinfo {author} {\bibfnamefont {E.}~\bibnamefont
  {Levin}},\ }\bibfield  {title} {\bibinfo {title} {{High energy QCD:
  multiplicity distribution and entanglement entropy}},\ }\href
  {https://doi.org/10.1103/PhysRevD.102.074008} {\bibfield  {journal} {\bibinfo
   {journal} {Phys. Rev. D}\ }\textbf {\bibinfo {volume} {102}},\ \bibinfo
  {pages} {074008} (\bibinfo {year} {2020})},\ \Eprint
  {https://arxiv.org/abs/2006.11793} {arXiv:2006.11793 [hep-ph]} \BibitemShut
  {NoStop}%
\bibitem [{\citenamefont {Koba}\ \emph {et~al.}(1972)\citenamefont {Koba},
  \citenamefont {Nielsen},\ and\ \citenamefont {Olesen}}]{Koba:1972ng}%
  \BibitemOpen
  \bibfield  {author} {\bibinfo {author} {\bibfnamefont {Z.}~\bibnamefont
  {Koba}}, \bibinfo {author} {\bibfnamefont {H.~B.}\ \bibnamefont {Nielsen}},\
  and\ \bibinfo {author} {\bibfnamefont {P.}~\bibnamefont {Olesen}},\
  }\bibfield  {title} {\bibinfo {title} {{Scaling of multiplicity distributions
  in high-energy hadron collisions}},\ }\href
  {https://doi.org/10.1016/0550-3213(72)90551-2} {\bibfield  {journal}
  {\bibinfo  {journal} {Nucl. Phys. B}\ }\textbf {\bibinfo {volume} {40}},\
  \bibinfo {pages} {317} (\bibinfo {year} {1972})}\BibitemShut {NoStop}%
\bibitem [{\citenamefont {Grosse-Oetringhaus}\ and\ \citenamefont
  {Reygers}(2010)}]{Grosse-Oetringhaus:2009eis}%
  \BibitemOpen
  \bibfield  {author} {\bibinfo {author} {\bibfnamefont {J.~F.}\ \bibnamefont
  {Grosse-Oetringhaus}}\ and\ \bibinfo {author} {\bibfnamefont
  {K.}~\bibnamefont {Reygers}},\ }\bibfield  {title} {\bibinfo {title}
  {{Charged-particle multiplicity in proton-proton collisions}},\ }\href
  {https://doi.org/10.1088/0954-3899/37/8/083001} {\bibfield  {journal}
  {\bibinfo  {journal} {J. Phys. G}\ }\textbf {\bibinfo {volume} {37}},\
  \bibinfo {pages} {083001} (\bibinfo {year} {2010})},\ \Eprint
  {https://arxiv.org/abs/0912.0023} {arXiv:0912.0023 [hep-ex]} \BibitemShut
  {NoStop}%
\bibitem [{\citenamefont {Levin}\ and\ \citenamefont
  {Lublinsky}(2004)}]{Levin:2003nc}%
  \BibitemOpen
  \bibfield  {author} {\bibinfo {author} {\bibfnamefont {E.}~\bibnamefont
  {Levin}}\ and\ \bibinfo {author} {\bibfnamefont {M.}~\bibnamefont
  {Lublinsky}},\ }\bibfield  {title} {\bibinfo {title} {{A linear evolution for
  non-linear dynamics and correlations in realistic nuclei}},\ }\href
  {https://doi.org/10.1016/j.nuclphysa.2003.10.020} {\bibfield  {journal}
  {\bibinfo  {journal} {Nucl. Phys. A}\ }\textbf {\bibinfo {volume} {730}},\
  \bibinfo {pages} {191} (\bibinfo {year} {2004})},\ \Eprint
  {https://arxiv.org/abs/hep-ph/0308279} {arXiv:hep-ph/0308279} \BibitemShut
  {NoStop}%
\bibitem [{\citenamefont {Stoffers}\ and\ \citenamefont
  {Zahed}(2013)}]{Stoffers:2012mn}%
  \BibitemOpen
  \bibfield  {author} {\bibinfo {author} {\bibfnamefont {A.}~\bibnamefont
  {Stoffers}}\ and\ \bibinfo {author} {\bibfnamefont {I.}~\bibnamefont
  {Zahed}},\ }\bibfield  {title} {\bibinfo {title} {{Holographic Pomeron and
  entropy}},\ }\href {https://doi.org/10.1103/PhysRevD.88.025038} {\bibfield
  {journal} {\bibinfo  {journal} {Phys. Rev. D}\ }\textbf {\bibinfo {volume}
  {88}},\ \bibinfo {pages} {025038} (\bibinfo {year} {2013})},\ \Eprint
  {https://arxiv.org/abs/1211.3077} {arXiv:1211.3077 [nucl-th]} \BibitemShut
  {NoStop}%
\bibitem [{\citenamefont {Germano}\ and\ \citenamefont
  {Navarra}(2022)}]{Germano:2021brq}%
  \BibitemOpen
  \bibfield  {author} {\bibinfo {author} {\bibfnamefont {G.~R.}\ \bibnamefont
  {Germano}}\ and\ \bibinfo {author} {\bibfnamefont {F.~S.}\ \bibnamefont
  {Navarra}},\ }\bibfield  {title} {\bibinfo {title} {{Energy dependence of the
  multiplicity moments at the LHC}},\ }\href
  {https://doi.org/10.1103/PhysRevD.105.014005} {\bibfield  {journal} {\bibinfo
   {journal} {Phys. Rev. D}\ }\textbf {\bibinfo {volume} {105}},\ \bibinfo
  {pages} {014005} (\bibinfo {year} {2022})},\ \Eprint
  {https://arxiv.org/abs/2110.12028} {arXiv:2110.12028 [hep-ph]} \BibitemShut
  {NoStop}%
\bibitem [{\citenamefont {Bartels}\ \emph {et~al.}(2003)\citenamefont
  {Bartels}, \citenamefont {Gotsman}, \citenamefont {Levin}, \citenamefont
  {Lublinsky},\ and\ \citenamefont {Maor}}]{Bartels:2002uf}%
  \BibitemOpen
  \bibfield  {author} {\bibinfo {author} {\bibfnamefont {J.}~\bibnamefont
  {Bartels}}, \bibinfo {author} {\bibfnamefont {E.}~\bibnamefont {Gotsman}},
  \bibinfo {author} {\bibfnamefont {E.}~\bibnamefont {Levin}}, \bibinfo
  {author} {\bibfnamefont {M.}~\bibnamefont {Lublinsky}},\ and\ \bibinfo
  {author} {\bibfnamefont {U.}~\bibnamefont {Maor}},\ }\bibfield  {title}
  {\bibinfo {title} {{The dipole picture and saturation in soft processes}},\
  }\href {https://doi.org/10.1016/S0370-2693(03)00128-X} {\bibfield  {journal}
  {\bibinfo  {journal} {Phys. Lett. B}\ }\textbf {\bibinfo {volume} {556}},\
  \bibinfo {pages} {114} (\bibinfo {year} {2003})},\ \Eprint
  {https://arxiv.org/abs/hep-ph/0212284} {arXiv:hep-ph/0212284} \BibitemShut
  {NoStop}%
\bibitem [{\citenamefont {Carvalho}\ \emph {et~al.}(2008)\citenamefont
  {Carvalho}, \citenamefont {Duraes}, \citenamefont {Goncalves},\ and\
  \citenamefont {Navarra}}]{Carvalho:2007cf}%
  \BibitemOpen
  \bibfield  {author} {\bibinfo {author} {\bibfnamefont {F.}~\bibnamefont
  {Carvalho}}, \bibinfo {author} {\bibfnamefont {F.~O.}\ \bibnamefont
  {Duraes}}, \bibinfo {author} {\bibfnamefont {V.~P.}\ \bibnamefont
  {Goncalves}},\ and\ \bibinfo {author} {\bibfnamefont {F.~S.}\ \bibnamefont
  {Navarra}},\ }\bibfield  {title} {\bibinfo {title} {{Gluon saturation and the
  Froissart bound: A Simple approach}},\ }\href
  {https://doi.org/10.1142/S0217732308028417} {\bibfield  {journal} {\bibinfo
  {journal} {Mod. Phys. Lett. A}\ }\textbf {\bibinfo {volume} {23}},\ \bibinfo
  {pages} {2847} (\bibinfo {year} {2008})},\ \Eprint
  {https://arxiv.org/abs/0705.1842} {arXiv:0705.1842 [hep-ph]} \BibitemShut
  {NoStop}%
\bibitem [{\citenamefont {Khachatryan}\ \emph {et~al.}(2011)\citenamefont
  {Khachatryan} \emph {et~al.}}]{CMS:2010qvf}%
  \BibitemOpen
  \bibfield  {author} {\bibinfo {author} {\bibfnamefont {V.}~\bibnamefont
  {Khachatryan}} \emph {et~al.} (\bibinfo {collaboration} {CMS}),\ }\bibfield
  {title} {\bibinfo {title} {{Charged particle multiplicities in $pp$
  interactions at $\sqrt{s}=0.9$, 2.36, and 7 TeV}},\ }\href
  {https://doi.org/10.1007/JHEP01(2011)079} {\bibfield  {journal} {\bibinfo
  {journal} {J. High Energy Phys.}\ }\textbf {\bibinfo {volume} {01}},\
  \bibinfo {pages} {079}},\ \Eprint {https://arxiv.org/abs/1011.5531}
  {arXiv:1011.5531 [hep-ex]} \BibitemShut {NoStop}%
\bibitem [{\citenamefont {Acharya}\ \emph {et~al.}(2023)\citenamefont {Acharya}
  \emph {et~al.}}]{ALICE:2022xip}%
  \BibitemOpen
  \bibfield  {author} {\bibinfo {author} {\bibfnamefont {S.}~\bibnamefont
  {Acharya}} \emph {et~al.} (\bibinfo {collaboration} {ALICE}),\ }\bibfield
  {title} {\bibinfo {title} {{Multiplicity dependence of charged-particle
  production in pp, p-Pb, Xe-Xe and Pb-Pb collisions at the LHC}},\ }\href
  {https://doi.org/10.1016/j.physletb.2023.138110} {\bibfield  {journal}
  {\bibinfo  {journal} {Phys. Lett. B}\ }\textbf {\bibinfo {volume} {845}},\
  \bibinfo {pages} {138110} (\bibinfo {year} {2023})},\ \Eprint
  {https://arxiv.org/abs/2211.15326} {arXiv:2211.15326 [nucl-ex]} \BibitemShut
  {NoStop}%
\bibitem [{\citenamefont {Adloff}\ \emph {et~al.}(1996)\citenamefont {Adloff}
  \emph {et~al.}}]{H1:1996naa}%
  \BibitemOpen
  \bibfield  {author} {\bibinfo {author} {\bibfnamefont {C.}~\bibnamefont
  {Adloff}} \emph {et~al.} (\bibinfo {collaboration} {H1}),\ }\bibfield
  {title} {\bibinfo {title} {{Inclusive $D^0$ and $D^{* \pm}$ production in
  deep inelastic $ep$ scattering at HERA}},\ }\href
  {https://doi.org/10.1007/s002880050281} {\bibfield  {journal} {\bibinfo
  {journal} {Z. Phys. C}\ }\textbf {\bibinfo {volume} {72}},\ \bibinfo {pages}
  {593} (\bibinfo {year} {1996})},\ \Eprint
  {https://arxiv.org/abs/hep-ex/9607012} {arXiv:hep-ex/9607012} \BibitemShut
  {NoStop}%
\bibitem [{\citenamefont {Breitweg}\ \emph {et~al.}(2000)\citenamefont
  {Breitweg} \emph {et~al.}}]{ZEUS:1999qnm}%
  \BibitemOpen
  \bibfield  {author} {\bibinfo {author} {\bibfnamefont {J.}~\bibnamefont
  {Breitweg}} \emph {et~al.} (\bibinfo {collaboration} {ZEUS}),\ }\bibfield
  {title} {\bibinfo {title} {{Measurement of $D^{* \pm}$ production and the
  charm contribution to $F_2$ in deep inelastic scattering at HERA}},\ }\href
  {https://doi.org/10.1007/s100529900244} {\bibfield  {journal} {\bibinfo
  {journal} {Eur. Phys. J. C}\ }\textbf {\bibinfo {volume} {12}},\ \bibinfo
  {pages} {35} (\bibinfo {year} {2000})},\ \Eprint
  {https://arxiv.org/abs/hep-ex/9908012} {arXiv:hep-ex/9908012} \BibitemShut
  {NoStop}%
\bibitem [{\citenamefont {Ademollo}\ and\ \citenamefont
  {Del~Giudice}(1969)}]{AG69}%
  \BibitemOpen
  \bibfield  {author} {\bibinfo {author} {\bibfnamefont {M.}~\bibnamefont
  {Ademollo}}\ and\ \bibinfo {author} {\bibfnamefont {E.}~\bibnamefont
  {Del~Giudice}},\ }\bibfield  {title} {\bibinfo {title} {{Nonstrong amplitudes
  in a Veneziano-type model}},\ }\href {https://doi.org/10.1007/BF02756239}
  {\bibfield  {journal} {\bibinfo  {journal} {Nuovo Cim. A}\ }\textbf {\bibinfo
  {volume} {63}},\ \bibinfo {pages} {639} (\bibinfo {year} {1969})}\BibitemShut
  {NoStop}%
\bibitem [{\citenamefont {Landshoff}\ and\ \citenamefont
  {Polkinghorne}(1970)}]{LP70}%
  \BibitemOpen
  \bibfield  {author} {\bibinfo {author} {\bibfnamefont {P.~V.}\ \bibnamefont
  {Landshoff}}\ and\ \bibinfo {author} {\bibfnamefont {J.~C.}\ \bibnamefont
  {Polkinghorne}},\ }\bibfield  {title} {\bibinfo {title} {{The scaling law for
  deep inelastic scattering in a new Veneziano-like amplitude}},\ }\href
  {https://doi.org/10.1016/0550-3213(70)90359-7} {\bibfield  {journal}
  {\bibinfo  {journal} {Nucl. Phys. B}\ }\textbf {\bibinfo {volume} {19}},\
  \bibinfo {pages} {432} (\bibinfo {year} {1970})}\BibitemShut {NoStop}%
\bibitem [{\citenamefont {Bekenstein}(1972)}]{Bekenstein:1972tm}%
  \BibitemOpen
  \bibfield  {author} {\bibinfo {author} {\bibfnamefont {J.~D.}\ \bibnamefont
  {Bekenstein}},\ }\bibfield  {title} {\bibinfo {title} {{Black holes and the
  second law}},\ }\href {https://doi.org/10.1007/BF02757029} {\bibfield
  {journal} {\bibinfo  {journal} {Lett. Nuovo Cim.}\ }\textbf {\bibinfo
  {volume} {4}},\ \bibinfo {pages} {737} (\bibinfo {year} {1972})}\BibitemShut
  {NoStop}%
\bibitem [{\citenamefont {Hawking}(1975)}]{Hawking:1975vcx}%
  \BibitemOpen
  \bibfield  {author} {\bibinfo {author} {\bibfnamefont {S.~W.}\ \bibnamefont
  {Hawking}},\ }\bibfield  {title} {\bibinfo {title} {{Particle creation by
  black holes}},\ }\href {https://doi.org/10.1007/BF02345020} {\bibfield
  {journal} {\bibinfo  {journal} {Commun. Math. Phys.}\ }\textbf {\bibinfo
  {volume} {43}},\ \bibinfo {pages} {199} (\bibinfo {year} {1975})},\ \bibinfo
  {note} {[Erratum: Commun. Math. Phys. 46, 206 (1976)]}\BibitemShut {NoStop}%
\bibitem [{\citenamefont {Witten}(2001)}]{Witten:2001}%
  \BibitemOpen
  \bibfield  {author} {\bibinfo {author} {\bibfnamefont {E.}~\bibnamefont
  {Witten}},\ }\href@noop {} {\bibinfo {title} {{Black holes and quark
  confinement}}},\ \bibinfo {howpublished} {Available at
  \url{https://www.ias.edu/sites/default/files/sns/CurrentScienceVol81(4).pdf}}
  (\bibinfo {year} {2001})\BibitemShut {NoStop}%
\end{thebibliography}%

\end{document}